\documentclass[12pt]{article}

\usepackage{verbatim,color,amssymb}
\usepackage{amsmath}					
\usepackage{amsthm}					
\usepackage{natbib}
\usepackage{multirow}
\usepackage{setspace}
\usepackage[mathscr]{euscript}
\usepackage{fancyhdr}
\usepackage{enumitem}
\usepackage{graphicx}
\usepackage{geometry}
\usepackage[bookmarks=false]{hyperref}
\usepackage{dsfont}
\usepackage[table]{xcolor}
\usepackage{tikz}
\usetikzlibrary{arrows}
\usetikzlibrary{patterns}
\usepackage{algorithm, algorithmicx}
\usepackage{multibib}	
\newcites{latex}{Additional References}

\usepackage{tabularx, booktabs}
\newcolumntype{Y}{>{\centering\arraybackslash}X}
\usepackage{lscape}

\usepackage{caption}
\usepackage{subcaption}
\usepackage{float}
\usepackage{lineno}

\newtheorem{Thm}{\underline{\bf Theorem}}

\newtheorem{Proof}{Proof}
\newtheorem*{Proof*}{Proof}

\newtheorem{Lem}{\underline{\bf Lemma}}

\newcommand{\mytilde}[2][1.2]{
  \mathrel{\overset{\mbox{\tiny #2}}{\scalebox{#1}[1]{$\sim$}}}}

\def\log{\hbox{log}}

\def\Beta{\hbox{Beta}}

\def\Dir{\hbox{Dir}}

\def\Exp{\hbox{Exp}}

\def\IG{\hbox{Inv-Ga}}

\def\Normal{\hbox{Normal}}

\def\Unif{\hbox{Unif}}

\def\P_25_ICML{{\it Proceedings of the 25th international conference on Machine learning}}

\def\bse{\begin{eqnarray*}}
\def\ese{\end{eqnarray*}}
\def\be{\begin{eqnarray}}
\def\ee{\end{eqnarray}}
\def\bq{\begin{equation}}
\def\eq{\end{equation}}

\def\bC{{\mathbf C}}
\def\bd{{\mathbf d}}

\def\b1e{{\mathbf e}}
\def\b1f{{\mathbf f}}

\def\bm{{\mathbf m}}

\def\bp{{\mathbf p}}

\def\br{{\mathbf r}}

\def\bs{{\mathbf s}}

\def\bx{{\mathbf x}}

\def\bY{{\mathbf Y}}
\def\bz{{\mathbf z}}

\newcommand{\bmu}{\mbox{\boldmath $\mu$}}

\newcommand{\bpi}{\mbox{\boldmath $\pi$}}

\newcommand{\btheta}{\mbox{\boldmath $\theta$}}

\newcommand{\bzeta}{\mbox{\boldmath $\zeta$}}
\newcommand{\bsigma}{\mbox{\boldmath $\sigma$}}

\renewcommand\footnoterule{\kern-3pt \hrule \textwidth 2in \kern 2.6pt}

\newcommand\Algphase[1]{%
\vspace*{-.5\baselineskip}\Statex\hspace*{\dimexpr-\algorithmicindent-2pt\relax}\rule{\textwidth}{0.4pt}%
\Statex\hspace*{-\algorithmicindent}\textbf{#1}%
\vspace*{-.5\baselineskip}\Statex\hspace*{\dimexpr-\algorithmicindent-2pt\relax}\rule{\textwidth}{0.4pt}%
}

\def\boxit#1{\vbox{\hrule\hbox{\vrule\kern6pt \vbox{\kern6pt \textcolor{blue}{#1}\kern6pt}\kern6pt\vrule}\hrule}}

\def\authorfootnote#1{{\let\thefootnote\relax\footnotetext{#1}}}

\definecolor{indianred3}{RGB}{205, 85, 85}

\allowdisplaybreaks

\begin{document}
\thispagestyle{empty}
\baselineskip=28pt

\begin{center}
{\LARGE{\bf Bayesian Nonparametric \\ 
\vskip -8pt Bivariate Survival Regression \\
for Current Status Data}}
\end{center}
\baselineskip=12pt
\vskip 20pt

\begin{center}
Giorgio Paulon$^{1}$ (giorgio.paulon@utexas.edu)\\
Peter M\"uller$^{2}$ (pmueller@math.utexas.edu)\\
Victor G. Sal Y Rosas$^{3}$(vsalyrosas@pucp.edu.pe)\\

\vskip 7mm
$^{1}$Department of Statistics and Data Sciences, \\
University of Texas at Austin,\\ 2317 Speedway D9800, Austin, TX 78712-1823, USA\\
\vskip 8pt 
$^{2}$Department of Mathematics, \\
University of Texas at Austin,\\ 2515 Speedway C1200, Austin, TX 78712-1202, USA\\
\vskip 8pt 
$^{3}$Secci\'{o}n Matem\'{a}ticas, Departamento de Ciencias,\\ 
Pontificia Universidad Cat\'{o}lica del Per\'{u},\\
Av. Universitaria 1801, San Miguel 15088, Peru
\end{center}

\vskip 20pt 
\begin{center}
{\Large{\bf Abstract}} 
\end{center}
\baselineskip=12pt

We consider nonparametric inference for event time distributions based on current status data.
We show that in this scenario conventional mixture priors, including the popular Dirichlet process mixture prior, lead to biologically uninterpretable results as they unnaturally skew the probability mass for the event times toward the extremes of the observed data.
Simple assumptions on dependent censoring can fix the problem. 
We then extend the discussion to bivariate current status data with partial ordering of the two outcomes.
In addition to dependent censoring, we also exploit some minimal known structure relating the two event times. 
We design a Markov chain Monte Carlo algorithm for posterior simulation. 
Applied to a recurrent infection study, the method provides novel insights into how symptoms-related hospital visits are affected by covariates.

\vskip 20pt 
\baselineskip=12pt
\noindent\underline{\bf Key Words}: 
Survival regression;
Current status data;
Bayesian nonparametrics;
Joint modeling;
Race model;
Recurrent infections

\par\medskip\noindent
\underline{\bf Short/Running Title}: Bivariate Survival Regression

\par\medskip\noindent
\underline{\bf Corresponding Author}: Giorgio Paulon (giorgio.paulon@utexas.edu)

\pagenumbering{arabic}
\setcounter{page}{0}
\newlength{\gnat}
\setlength{\gnat}{16pt}
\baselineskip=\gnat

\newpage 

\section{Introduction}

We develop Bayesian nonparametric survival regression for bivariate event times that are subject to a single censoring time.
In particular, we consider bivariate current status data \citep{groeneboom1992information}, referring to situations where the only available information on each event time is whether or not it exceeds a monitoring time that is common to the two outcomes.
Data of this type are often collected in studies on the prevalence of recurrent infectious diseases.  
We show that standard survival analysis models \citep{ibrahim2005b} fail to provide a meaningful estimate of the latent event time distribution when applied to current status data. 
The analysis of this kind of data is complicated by the fact that all event times are either left or right censored.
We propose a modeling approach that addresses this gap in the literature.

Our goal is to develop a flexible model whose parameters have a biologically meaningful interpretation. 
Bayesian models are especially useful in such scenarios because of their ability to accommodate prior information.
Nonparametric priors are often used to flexibly model a baseline survival function, usually completed with a parametric component that relates survival to a number of predictors. 
For example, extensions of the proportional hazards (PH) model \citep{cox1972regression} have been proposed in \citet{kalbfleisch1978non} and in \citet{hjort1990nonparametric}.
Generalizations of the accelerated failure times (AFT) model \citep{buckley1979linear} based on a Dirichlet process prior appear in \citet{christensen1988modelling}, \citet{kuo1997bayesian}, \citet{kottas2001bayesian}, \citet{hanson2004bayesian}, or alternatively using Polya trees, for example in \citet{hanson2002modeling}.
In other cases the main inference target is the hazard function.
\citet{sparapani2016nonparametric}, for instance, construct nonparametric survival regression using a Bayesian additive regression tree (BART) model \citep{chipman2010bart} by adding time as an ordinal predictor to a BART-probit model for the hazard function.

In general, censored observations contribute limited information, via the distribution function or survival function as the corresponding factors of the joint likelihood. 
This becomes problematic in the case of current status data, as we shall demonstrate. 
Some proposals have been put forward to tackle these issues.
In the case of survival regression, generalizations of the PH model for current status data have been introduced in \citet{cai2011bayesian} and in \citet{wang2015regression}, focusing on the univariate case. 
More similar to our approach, \citet{wang2000assessing} model dependence between bivariate event times via a copula function.
\citet{dunson2002bayesian} use a Bayesian probit model with normal frailties to induce dependence among multivariate current status data.
Nevertheless, there remains a gap in the literature concerning fully nonparametric regression for bivariate current status data with dependent censoring.

The motivating case study is inference for the Partner Notification Study \citep{golden2005effect}.
The goal of the study is to understand the times of development of infection and symptoms for recurrent episodes of gonorrhea and/or chlamydial infections.
The study design includes a single follow-up visit for each individual. 
During this visit the presence of symptoms and infection was recorded, leading to all censored data with shared censoring times for the two outcomes.

Let $S$ denote the time of the onset of symptoms, $I$ the time of infection, and $C$ the time of the hospital visit. 
Thus, four responses are possible: presence of both disease and symptoms ($I < C,S < C$), absence of both ($I > C,S > C$), absence of symptoms and presence of disease ($I < C, S > C$), and symptoms without disease ($I > C, S < C$).
The latter can be explained by the fact that the surveyed symptoms are very generic and might also arise due to other underlying causes. 
This setup yields data that are bivariate in nature as two outcomes are registered. 
However, the censoring times, i.e. the hospital visit times, are restricted to a lower dimensional subspace, with a single follow-up visit to assess the presence of both symptoms and disease.
Additional complexity arises from the partial ordering of the two outcomes: the infection time is a priori unlikely to follow the symptoms time.
This can only occur when the symptoms arise due to other causes. 
Our model introduces features to reflect this consideration. 
We use a mixture model with one submodel being subject to an order constraint, representing symptoms due to the infection of interest, and another submodel without such constraint, allowing for symptoms due to other causes. 
While our discussion is motivated by a specific application, we note that similar data formats arise frequently in any study that involves data collection during follow-up visits. 
For example, doctors might record tumor recurrence using a CT scan and symptoms as reported by patients.

In the first part of this article, we demonstrate with simple examples the problems arising from the use of standard techniques with current status data.
We then introduce structural assumptions that allow us to identify a meaningful distribution of the latent bivariate outcomes. 
We propose a Bayesian nonparametric (BNP) approach for modeling the joint distribution under these assumptions. 
An important feature of BNP models is their large support, allowing us to approximate essentially arbitrary distributions \citep{ishwaran2001gibbs}. 
To handle covariates, our approach is based on the dependent Dirichlet process (DDP) prior introduced by \citet{maceachern1999dependent}.
See also the discussion in \citet{de2004anova} for the special case of categorical covariates.

The rest of this article is organized as follows. 
Section \ref{sec:real_Study} describes the clinical study that motivates this article.
Section \ref{sec:univariate} develops the proposed inference approach starting from a simple univariate case.
Section \ref{sec:bivariate} uses the univariate model as a building block for bivariate outcomes. 
Section \ref{sec:post_inference} outlines computational challenges and an MCMC strategy.
Section \ref{sec:application} presents the results of the proposed method applied to the Partner Notification Study. 
Section \ref{sec:discussion} finishes with concluding remarks.
Additional details, including proofs, the MCMC scheme, convergence diagnostics and simulation studies are deferred to the supplementary materials.

\section{The Partner Notification Study}
\label{sec:real_Study}

The Partner Notification Study \citep{golden2005effect} enrolled men and women who received a diagnosis of gonorrhea or genital chlamydia at most $14$ days prior to enrollment. 
It was conducted in King County Seattle (Washington state, U.S.A.) from September $1998$ to March $2003$.
Researchers contacted clinicians who diagnosed and treated the infections to seek permission to contact their patients. 
To minimize the likelihood of reinfection before randomization, patients who could not be contacted within $14$ days after treatment were not eligible for the study, yielding a total of $n = 1864$ participants. 
The study was designed to gather current status data of recurrent gonorrhea or chlamydial infection in patients $3$ to $19$ weeks after randomization to standard (control group, $933$ individuals) or expedited partner therapy (intervention group, $931$ individuals).
The primary outcome was persistent or recurrent gonorrhea and/or chlamydial infection in the original participants within $90$ days after enrollment, although actual follow up times varied considerably ($19$ to $161$ days) due to difficulty contacting participants and scheduling follow-up visits.

When visiting the hospital, two outcomes were recorded for each patient: presence of an infection ($I_{i}$) and of symptoms ($S_{i}$). 
Thus, two latent event times $(I_{i}, S_{i})$ correspond to a common censoring time $C_{i}$, i.e. the time of the hospital visit. 
The data record for each patient $C_{i}$, and whether the patient has already experienced the infection $\Delta_{I_{i}} = \mathds{1}(I_i < C_i)$ and some symptoms $\Delta_{S_{i}} = \mathds{1}(S_i < C_i)$. 
While in general symptoms should follow the onset of infection, the definition of symptoms in this study is very generic and they might also be due to other causes.
In the case $I_i < S_i$ it is impossible to tell whether symptoms are due to the disease of interest or any other cause, while when $I_i > S_i$ the symptoms are known to be due some other cause.

The recorded $n = 1832$ follow-up visits included patients reporting all four possible combinations of censoring for the two outcomes: $n_{00} = 1303$ patients did not experience symptoms and tested negative for the infection; $n_{10} = 121$ patients tested positive for the infection but were not experiencing any symptoms (asymptomatic infections); $n_{01} = 325$ patients tested negative for the infection but were experiencing symptoms (due to other causes); $n_{11} = 83$ patients tested positive for the infection and were also experiencing symptoms (symptomatic infections).

\begin{figure}[!ht]
	\centering
	\includegraphics[width=.49\linewidth]{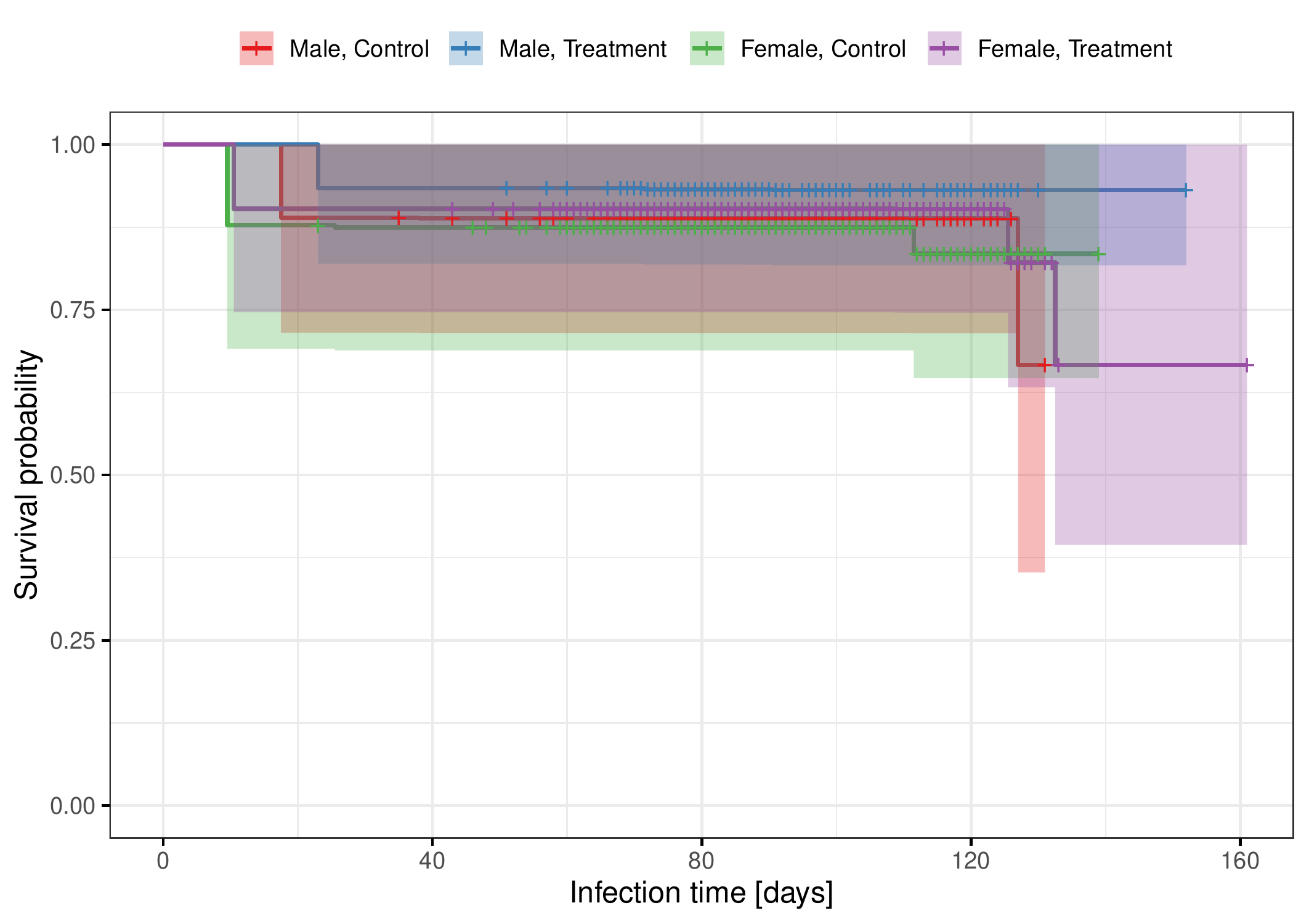}
	\includegraphics[width=.49\linewidth]{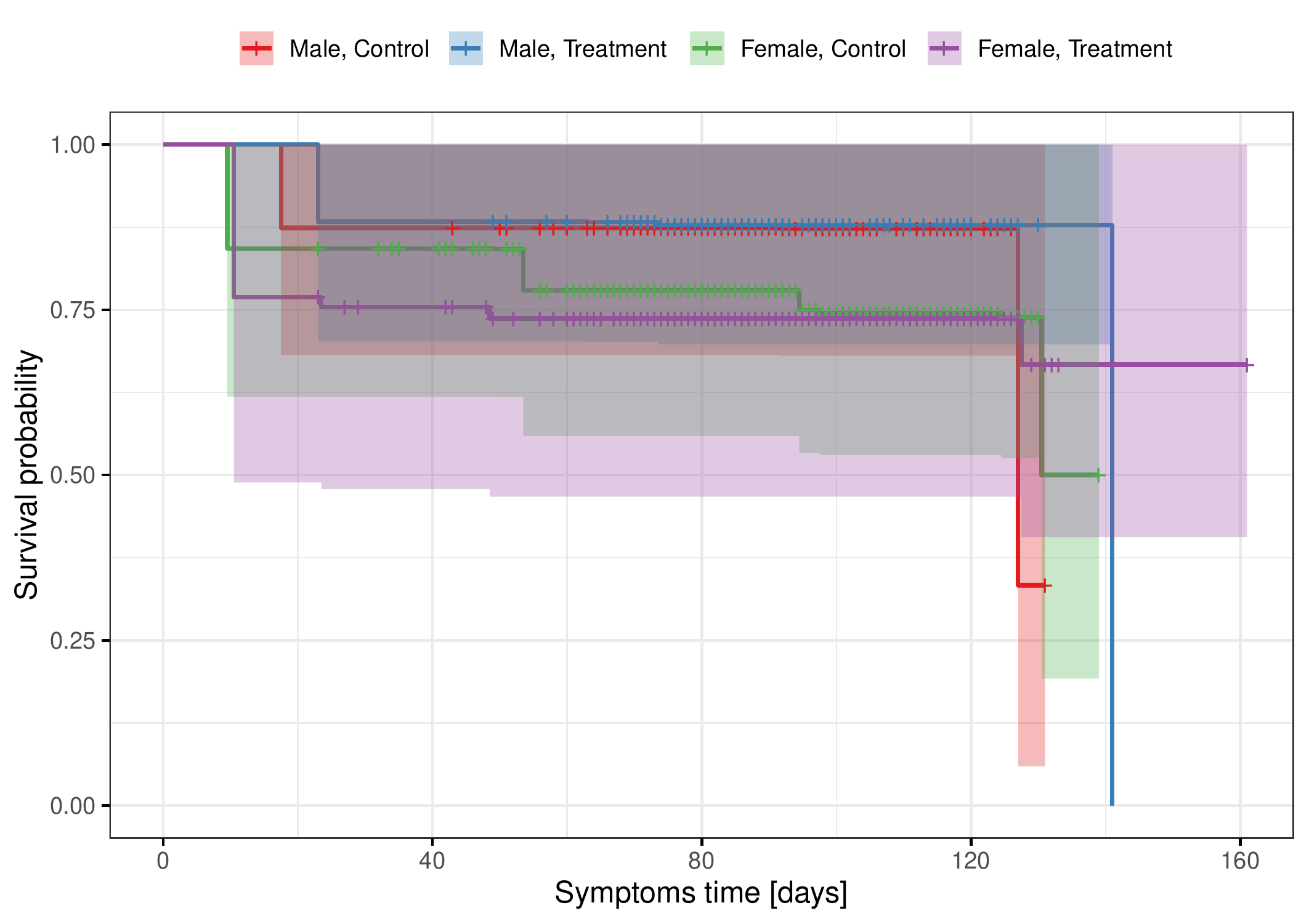}
	\caption{Nonparametric MLE estimate for infection times (left panel) and time until symptoms (right panel), stratified by the binary covariates \textit{gender} and \textit{treatment} fixing \textit{age} to the average age in the sample.}
	\label{fig:km_joint}
\end{figure}

Figure \ref{fig:km_joint} shows two univariate nonparametric maximum likelihood estimates (MLE) \citep{groeneboom1992information} for the distributions of time to infection $I_{i}$ and time to symptoms $S_{i}$, stratified by two covariates (gender and intervention). 
Female participants seem to experience symptoms sooner than men. 
The flat region of survival probability in the middle of the range of the observed data is due to the limited assumptions of the nonparametric MLE and is clinically highly implausible.
In Section \ref{sec:univariate} we show that the accumulation of probability mass toward the bounds of the observation range is a common issue when dealing with current status data.
Moreover, these nonparametric MLE estimates represent marginal effects and do not take into account any correlation that is expected between the time to infection and the time to development of symptoms.

\section{Univariate Survival Analysis for Current Status Data}
\label{sec:univariate}

We introduce a Bayesian nonparametric (BNP) modeling strategy for current status data, first in a simple univariate case. 
First, we show that the nonparametric MLE for current status data has an undesirable feature that makes it biologically uninterpretable. 
Most of the probability mass is accumulated toward the extremes of the data range.

Let $S_{i}$ represent the latent event time for patient $i$, $\Delta_{i}$ be a censoring indicator with $\Delta_{i} = 1$ if the event has been detected and $\Delta_{i} = 0$ otherwise , and let $C_{i}$ denote the censoring time. 
That is, when $\Delta_{i} = 1$, then $S_{i} \leq C_{i}$ (left censored), otherwise $S_{i} > C_{i}$ (right censored).
We want to infer the unknown distribution $f_{S}(s)$ based on only the observed censoring times and indicators $(C_{i}, \Delta_{i}), i = 1, \dots, n$.

\subsection{Limitations of the Maximum Likelihood Estimator}
\label{sec:EM}

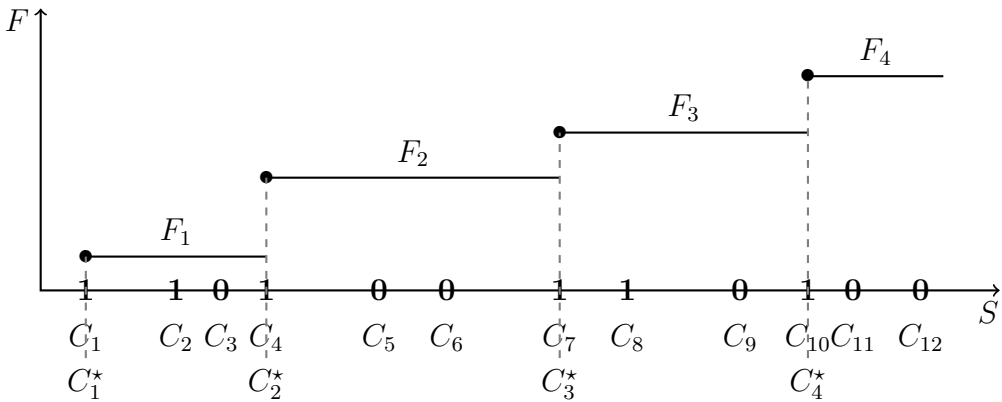
\begin{figure}[!ht]
	\begin{center}
	\begin{tikzpicture}[scale = 3]
		\draw [thick] (4.2,0) node[below]{$S$};		
		\draw [thick] (0,1.2) node[left]{$F$};
		\draw [<->, thick] (0,1.25) -- (0,0) -- (4.25,0);
		\draw [thick] (0.2,0.1) node[below]{\textbf{1}};
		\draw [thick] (0.6,0.1) node[below]{\textbf{1}};
		\draw [thick] (0.8,0.1) node[below]{\textbf{0}};
		\draw [thick] (1,0.1) node[below]{\textbf{1}};
		\draw [thick] (1.5,0.1) node[below]{\textbf{0}};
		\draw [thick] (1.8,0.1) node[below]{\textbf{0}};
		\draw [thick] (2.3,0.1) node[below]{\textbf{1}};
		\draw [thick] (2.6,0.1) node[below]{\textbf{1}};
		\draw [thick] (3.1,0.1) node[below]{\textbf{0}};
		\draw [thick] (3.4,0.1) node[below]{\textbf{1}};
		\draw [thick] (3.6,0.1) node[below]{\textbf{0}};
		\draw [thick] (3.9,0.1) node[below]{\textbf{0}};
		\draw [thick] (0.2,-0.1) node[below]{$C_1$};
		\draw [thick] (0.6,-0.1) node[below]{$C_2$};
		\draw [thick] (0.8,-0.1) node[below]{$C_3$};
		\draw [thick] (1,-0.1) node[below]{$C_4$};
		\draw [thick] (1.5,-0.1) node[below]{$C_5$};
		\draw [thick] (1.8,-0.1) node[below]{$C_6$};
		\draw [thick] (2.3,-0.1) node[below]{$C_7$};
		\draw [thick] (2.6,-0.1) node[below]{$C_8$};
		\draw [thick] (3.1,-0.1) node[below]{$C_9$};
		\draw [thick] (3.4,-0.1) node[below]{$C_{10}$};
		\draw [thick] (3.6,-0.1) node[below]{$C_{11}$};
		\draw [thick] (3.9,-0.1) node[below]{$C_{12}$};
		\draw [thick] (0.2,-0.3) node[below]{$C_1^\star$};
		\draw [thick] (1,-0.3) node[below]{$C_2^\star$};
		\draw [thick] (2.3,-0.3) node[below]{$C_3^\star$};
		\draw [thick] (3.4,-0.3) node[below]{$C_{4}^\star$};
		\draw[thick] (0.2,0.15) -- (1,0.15);
		\draw[thick] (1,0.5) -- (2.3,0.5);
		\draw[thick] (2.3,0.7) -- (3.4,0.7);
		\draw[thick] (3.4,0.95) -- (4,0.95);
        \foreach \Point in {(0.2,0.15), (1,0.5), (2.3,0.7), (3.4,0.95)}{
            \node at \Point {\textbullet};
        }
		\draw [thick] (0.6,0.15) node[above]{$F_1$};
		\draw [thick] (1.65,0.5) node[above]{$F_2$};
		\draw [thick] (2.85,0.7) node[above]{$F_3$};
		\draw [thick] (3.7,0.95) node[above]{$F_4$};

		\draw [thick,dashed,gray] (0.2,-0.3) -- (0.2,0.15);
		\draw [thick,dashed,gray] (1,-0.3) -- (1,0.5);
		\draw [thick, dashed,gray] (2.3,-0.3) -- (2.3,0.7);
		\draw [thick,dashed,gray] (3.4,-0.3) -- (3.4,0.95);

	\end{tikzpicture}
	\end{center}
	\caption{An example with $n = 12$ latent event times. The set of support points is $A = \{1, 4, 7, 10\}$. On the $x$-axis, $0$ and $1$ indicate the values of $\Delta_{i}$.}
	\label{fig:EM_alg}
\end{figure}

We show that under moderate sample sizes the nonparametric MLE does not provide meaningful estimates of the latent time distribution for current status data. 
Without loss of generality, we assume that the censoring times are ordered, $C_{i} \leq C_{i-1}$, and that $\Delta_{1} = 1, \Delta_{n} = 0$.
Define $A = \{i > 1 \text{ s.t. } \Delta_{i} = 1, \Delta_{i-1} = 0\} \cup \{1\}$ as the set of indices of left censored observations immediately following a right censored observation, i.e. the set of indices of the pairs $(\Delta_{i-1}, \Delta_{i}) = (0,1)$.
Next, let $J = |A|$ and $\bC^\star = (C_1^\star, \dots, C_J^\star) = (C_{i}, i \in A)$ denote the corresponding censoring times. 
See Figure \ref{fig:EM_alg} for an illustration.

Let $C_{J+1}^{\star}$ denote any point to the right of the last right censored observation. 
The times $\bC^{\star} \cup \{C_{J+1}^{\star}\}$ are the only points where probability mass can accumulate under the nonparametric MLE. 
In other words, the support of a discrete nonparametric density estimate for the latent event times can have probability mass only at the left censoring times. More specifically, at (i) the left censored observation in every ``01'' pair, (ii) the first left censored observation, and (iii) any point to the right of the last right censored observation. 
To see this, write the unknown distribution $f_{S}(\cdot)$ of the latent times $S_{i}$ as a discrete probability measure with atoms at the $C_j^\star$, i.e.
\begin{equation}
f_{S}(s) = \sum_{j=1}^{J+1} p_j \delta_{C_{j}^\star}.
\label{eq:EM_model}
\end{equation}
We denote with $F_j = \sum_{k \leq j} p_k$ the cumulative density function (c.d.f.) and with $\bar{F}_j = 1 - F_j$ the survival function at the support points. 
To see that the nonparametric MLE for $f_{S}(s)$ can only have support on the set $\bC^\star$, assume that $f_{S}(s)$ were to include any additional probability mass $p$ at $C_{i} \neq C_j^\star, j = 1,\dots,J$. 
Let $j^\star = \text{max}_j \{C_j^\star < C_{i} \}$ and $j^{\prime} = \text{min}_j \{C_j^\star > C_{i} \}$ denote the point mass in $\bC^\star$ closest to $C_{i}$ from the left and from the right, respectively. 
Then, if $\Delta_{i} = 1$ one could move the probability mass $p$ to $C_{j^{\star}}^{\star}$, and if $\Delta_{i} = 0$ one could move the probability mass $p$ to $C_{j^{\prime}}^{\star}$.
Either would leave the likelihood function unchanged.

\citet{groeneboom1992information} introduce a simple EM algorithm to estimate the unknown c.d.f for the latent times. 
Let $l_j = \#\{C_{i} \text{ s.t. } \Delta_{i} = 1, C_{j}^\star \leq C_{i} < C_{j+1}^\star \}$ and $r_j = \#\{C_{i} \text{ s.t. } \Delta_{i} = 0, C_{j}^\star < C_{i} \leq C_{j+1}^\star \}$ denote the runs of left and right censored observations, respectively. 
Let $\bY = \{(C_{i}, \Delta_{i})\}_{i=1}^{n}$ denote the data.
The log-likelihood function under model \eqref{eq:EM_model} is 
\begin{equation*}
    \begin{split}
    \ell(\bp; \bY) &= \sum_{i=1}^{n} \{ \delta_{1}(\Delta_{i}) \cdot \log  F(C_{i}) + \delta_{0}(\Delta_{i}) \cdot \log \bar{F}(C_{i})\} 
    \\
    &= \sum_{j=1}^{J} \{ l_j \log F_j + r_j \log \bar{F}_j \}.
    \end{split} 
\end{equation*}
If instead we knew the latent times $\bz = \{S_{i}\}_{i=1}^{n}$, we could use the full data log-likelihood $\ell (\bp, \bz) = \sum_{j=1}^{J} n_{j} \log  (p_j)$ where $n_j = \#\{S_{i} = C_{j}^\star \}$.
The expectation of this full data log-likelihood with respect to $\bz$ involves only $\mathbb{E}(n_{j} \mid \bp)$. 
This motivates an easy Expectation Maximization (EM) algorithm, shown in Algorithm \ref{algo: EM_alg}.

\begin{algorithm}
\caption{(Expectation Maximization)}
\label{algo: EM_alg}
\begin{algorithmic}[1]
\vspace{0.2cm}

\Algphase{E-step}
\State
For given $\bp = \bp^{(t)}$, evaluate the expectation of the $n_{j}$'s w.r.t. the latent censored event times.
This involves distributing $l_j$ to all $C_{j^\prime}^\star$, $j^\prime \leq j$ with weights $p_{j^\prime} / F_j$; and $r_j$ to all $C_{j^\prime}^\star$, $j^\prime > j$ with weights $p_{j^\prime} / \bar{F}_j$, i.e. 
$$\tilde{n}_j = \mathbb{E}_{\bs}[n_{j} \mid \bp] = \sum_{h \leq j} l_h p_j/F_h + \sum_{h > j} r_h p_j/\bar{F}_h.$$

\Algphase{M-step}
\State
Replacing the unknown $n_j$'s with their expectations $\tilde{n}_{j}$ makes the maximization (w.r.t $\bp$) of the expected (w.r.t $\bz$) full data log-likelihood $\ell (\bp, \bz)$ straightforward, yielding the update 
$$p^{(t+1)}_{j} = \tilde{n}_{j}/n.$$

\end{algorithmic}
\end{algorithm}
\begin{figure}
    \centering
    \begin{subfigure}[t]{0.47\textwidth}
        \centering
        \includegraphics[width=\textwidth]{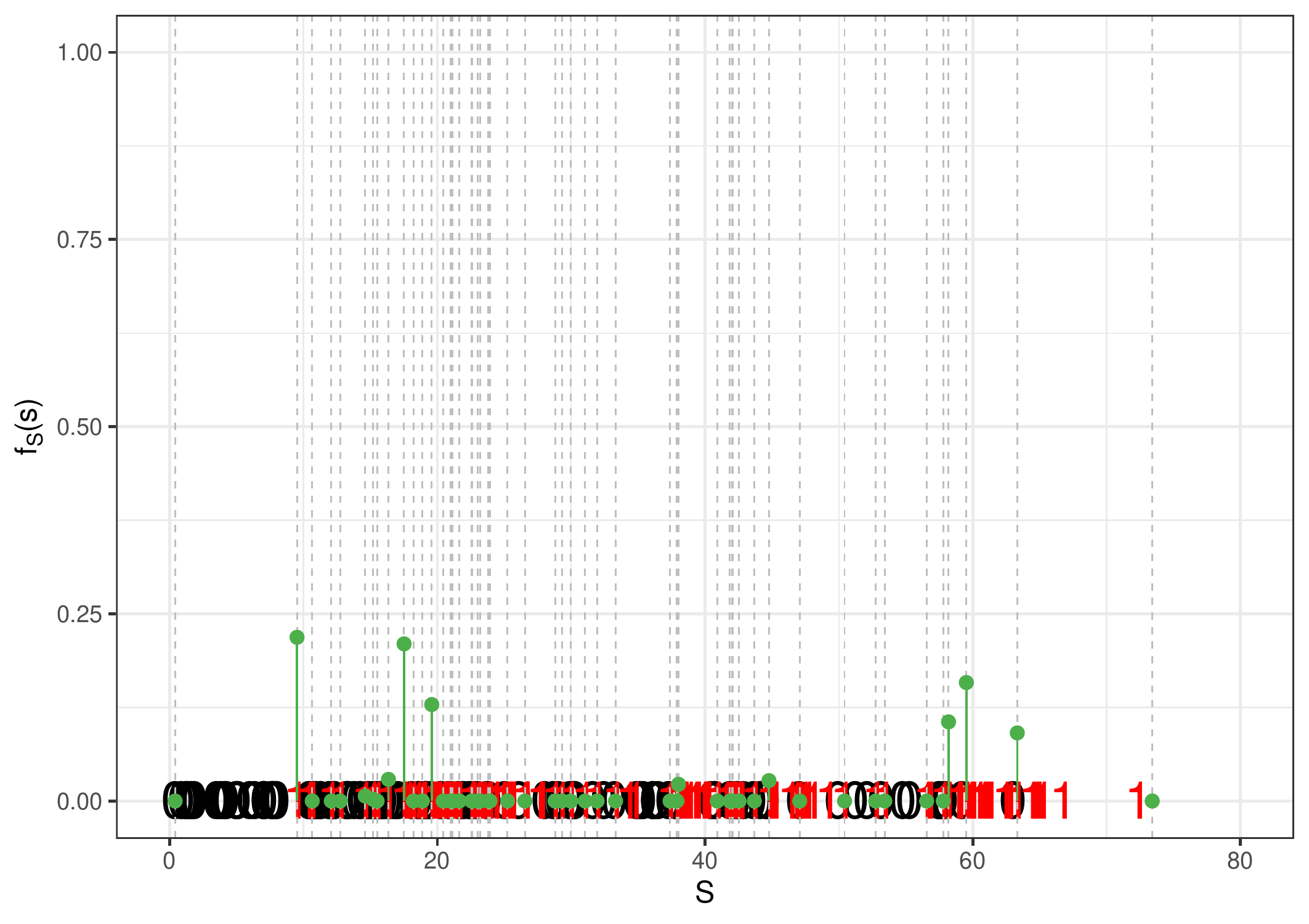}
        \caption{Vertical dashed lines represent the possible support points for $f_{S}(s)$. Green vertical pins represent the nonparametric MLE estimate for the point masses obtained via the EM algorithm.}
        \label{fig:em_result_1}
    \end{subfigure}
    \hfill
    \begin{subfigure}[t]{0.47\textwidth}
        \centering
        \includegraphics[width=\textwidth]{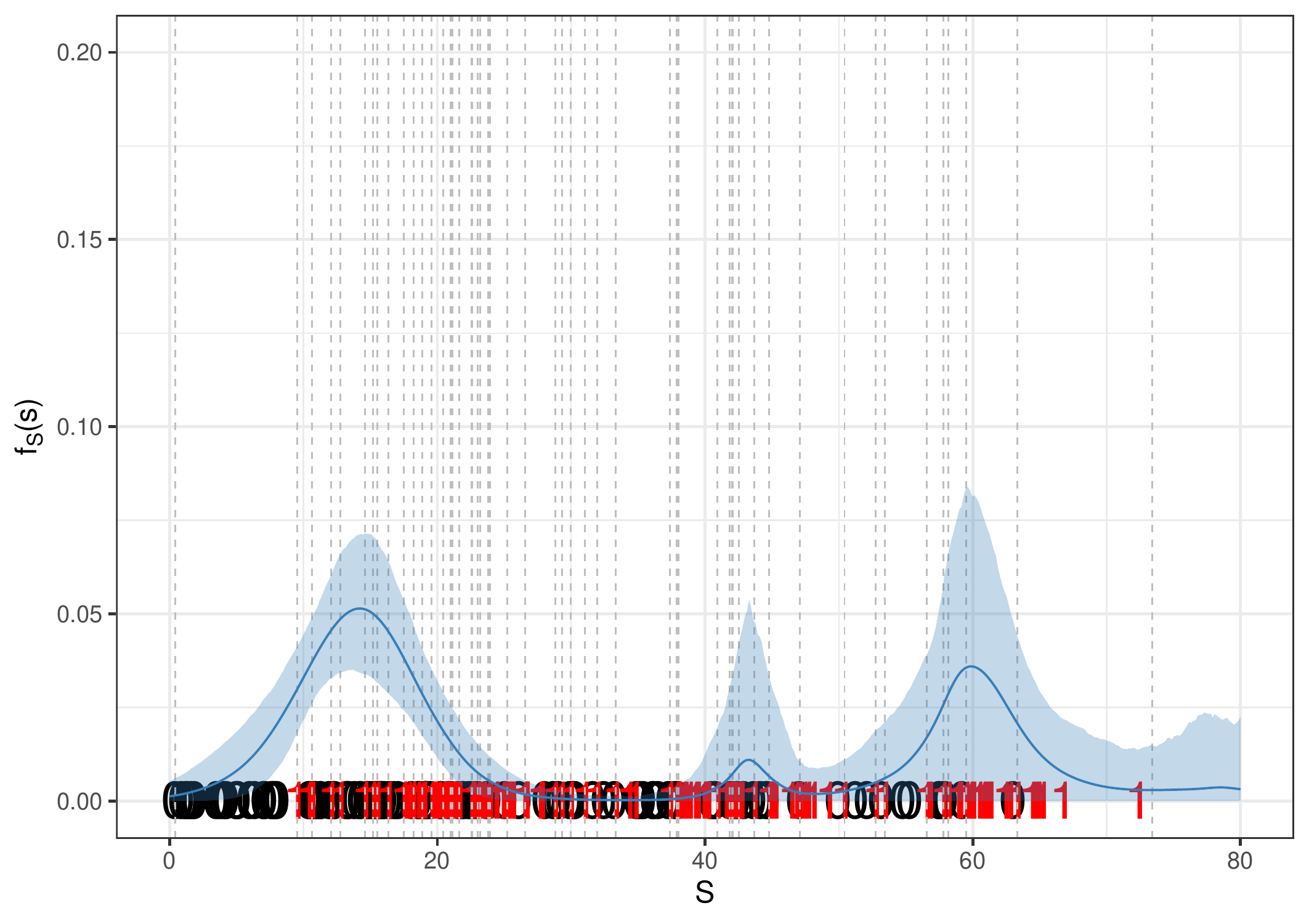}
        \caption{In blue, posterior mean for a simple mixture of $K = 3$ normal distributions. The shaded area represents the pointwise $95\%$ credible intervals for the posterior density estimate.}
        \label{fig:em_result_2}
    \end{subfigure}
    \caption{Simulated data. Right and left censoring times are represented by black ``0'' and red ``1'', respectively, on the $x$-axis.}
    \label{fig:em_result}
\end{figure}

We illustrate the algorithm on simulated data with $n = 200$ latent times generated from a mixture of three normal distributions with weights $\bpi = (0.4, 0.2, 0.4)^{\intercal}$, locations $\bmu = (20, 40, 60)^{\intercal}$ and scale parameters $\bsigma^{2} = (25, 25, 25)^{\intercal}$.
The censoring times $C_{i}$ were simulated according to model \eqref{eq:cens_model}, defined below.
As shown in Figure \ref{fig:em_result_1}, despite a large number of support points $\bC^{\star}$, in this simulation study most of the probability mass under the unconstrained MLE accumulates close to the bounds of the range of the data.
One might conjecture that the issue is caused by the excessively flexible nature of the unconstrained MLE.
However, even parametric models fail to capture the underlying distribution of the latent times. 
For comparison, we carried out inference using a mixture of $K = 3$ Gaussian distributions for the latent times $S$, matching the actual simulation truth. 
In Figure \ref{fig:em_result_2}, we show the posterior mean for the unknown event time distribution under this model when fitted to the current status data in the simulation study. 
The posterior estimated distribution still allocates most probability mass toward the extremes of the data, despite using an analysis model that matched the actual simulation truth.

\subsection{A Bayesian Nonparametric Model}
\label{sec:BNP_model}

We introduce some assumptions to address the issues described in the previous section. 
In short, we regularize the model by (i) explicitly modeling the dependence between censoring times and latent event times, and (ii) introducing prior shrinkage with a flexible Bayesian nonparametric prior.

Knowledge about dependent censoring allows us to gain some information on $f_{S}(\cdot)$ from the censoring times. 
For example, in the motivating case study it is expected that patients seek help shortly after they experience symptoms. 
This information can be incorporated in the model in many ways.
For our specific application, we assume that the censoring times $C_{i}$'s arise from a race between a return by schedule versus a return driven by the onset of symptoms, as
\begin{equation}
    C_{i} \mid S_{i}, \lambda = \min \{S_{i} + \Exp(\lambda); \Unif(A, B) \},
    \label{eq:cens_model}
\end{equation}
where $A$ and $B$ represent the range of the observation window, and $\Exp(\lambda)$ and $\Unif(A, B)$ refer to random variables with the respective distribution.
In other words, the visit time to the hospital can either occur uniformly in the observation range (visit by protocol) or it can closely follow the symptoms onset (visit prompted by symptoms). 
The resulting distribution is easily evaluated.
\begin{Lem}
The p.d.f. of the conditional distribution of censoring times given the event times is given by 
$$f_{C \mid S} (c \mid s) = \dfrac{\mathds{1}\{c \leq s\}}{B - A} + \dfrac{\mathds{1}\{c > s\}}{B - A} e^{-\lambda (c - s)} \{1 + \lambda (B - c)\}.$$
\end{Lem}

In addition to exploiting dependence, specifying a Bayesian nonparametric prior for the latent event time also helps to regularize inference on $f_{S}(\cdot)$. 
Relaxing parametric assumptions allows for greater modeling flexibility, robustness against misspecification of a parametric statistical model and, as a result, more honest uncertainty assessment.
At the same time, prior smoothing and shrinkage result in more realistic and clinically meaningful estimates.
In addition, a BNP model can allow to accommodate heterogeneous patient populations, for example using the Dirichlet process (DP) prior.
The DP was originally introduced by \citet{ferguson1973bayesian} and can be defined from its finite-dimensional analog. 
We write $H \sim DP(M, H_0)$ if the random distribution $H$ is such that for any partition $A_{1}, \dots, A_{K}$ of the sample space the random vector of the $H(A_{i})$ follows a Dirichlet distribution, $(H(A_{1}), \dots, H(A_{K})) \sim \Dir \{M H_{0}(A_{1}), \dots, M H_{0}(A_{K})\}$.
The DP prior is indexed by the total mass parameter $M$ (which controls the variance) and by the centering measure $H_{0}$ (which defines the expectation).
In fact, $\mathrm{E}[H(A)] = H_{0}(A)$ and
$\text{Var}[H(A)] = H_{0}(A) \{1 - H_{0}(A)\} / (M + 1)$.
Alternatively, \citet{sethuraman1994constructive} gives a constructive definition for the DP, known as the stick-breaking construction: $H = \sum_{k=1}^{+\infty} \pi_k \delta_{\theta_{k}}$ with $\pi_k = \nu_k \prod_{l < k} (1 - \nu_l), \nu_k \mytilde{iid} \text{Beta}(1, M)$ and $\theta_k \mytilde{iid} H_0$. 
In particular, the DP generates almost surely discrete probability measures. 
For this reason, often an additional convolution with a continuous kernel $k(y \mid \theta)$ is used to represent a random probability measure $F = \int k(y \mid \theta) dH(\theta) =  \sum_{k=1}^{+\infty} \pi_k k(y \mid \theta_{k})$ with $H \sim DP(M,H_{0})$. 
The model is known as DP mixture (DPM). 
Two natural choices of sampling models for survival data are the log normal and the Weibull families. 
In applications with event times close to $0$, it can be convenient to first log transform the data and then use normal kernels, i.e. use log normal kernels.
In many instances, however, a mixture of normals may suffice \citep{lo1984class} and is often preferred.

\subsubsection*{The BNP-CS model}
The resulting model can be summarized as follows
\begin{equation}
    \begin{split}
	    &C_{i} \mid S_i, \lambda = \min \{ S_{i} + \Exp(\lambda); \Unif(A, B) \}
	    \\
	    &S_i \mid H \sim \int \text{N}(S_i \mid \mu, \sigma^{2}) dH(\mu, \sigma^{2}), \quad H \sim \text{DP}(M, H_0).
    \end{split}
\label{eq:model_uni}
\end{equation}
The model is completed with hyperpriors, including
\begin{equation*}
    H_0 =  \text{N}(\mu_{k} \mid \mu_0, \sigma_{k}^{2}/\kappa_{0}) \times \text{IG}(\sigma_{k}^{2} \mid a_\sigma, b_\sigma),
\end{equation*}
$M \sim \text{Gamma} (a_{M}, b_{M})$ and $\lambda \sim \text{Gamma}(a_{\lambda}, b_{\lambda})$.
Using the stick-breaking construction of the DP, the second line of model \eqref{eq:model_uni} can be rewritten as 
$$S_i \mid \{\mu_k, \sigma_{k}^{2}, \pi_{k}\}_{k = 1}^{+\infty} \sim \sum_{k=1}^{+\infty} \pi_k \text{N}(S_i \mid \mu_k, \sigma^2_k)$$
with $( \mu_{k}, \sigma_{k}^{2} ) \sim H_0$, i.i.d., and $\bpi \sim \text{SB}(M)$, where $\text{SB}(M)$ denotes the stick-breaking construction for the weights, with concentration parameter $M$. 
In our implementation, we also use priors on the hyperparameters $\mu_{0}, \kappa_{0}, b_{\sigma}$.
We refer to \eqref{eq:model_uni} as BNP for current status (BNP-CS) model, with the name implying that alternative BNP priors other than the DPM \citep[see, e.g.][]{muller2015bayesian} could be used if desired.

\begin{figure}[!ht]
	\centering
	\includegraphics[width=.69\linewidth]{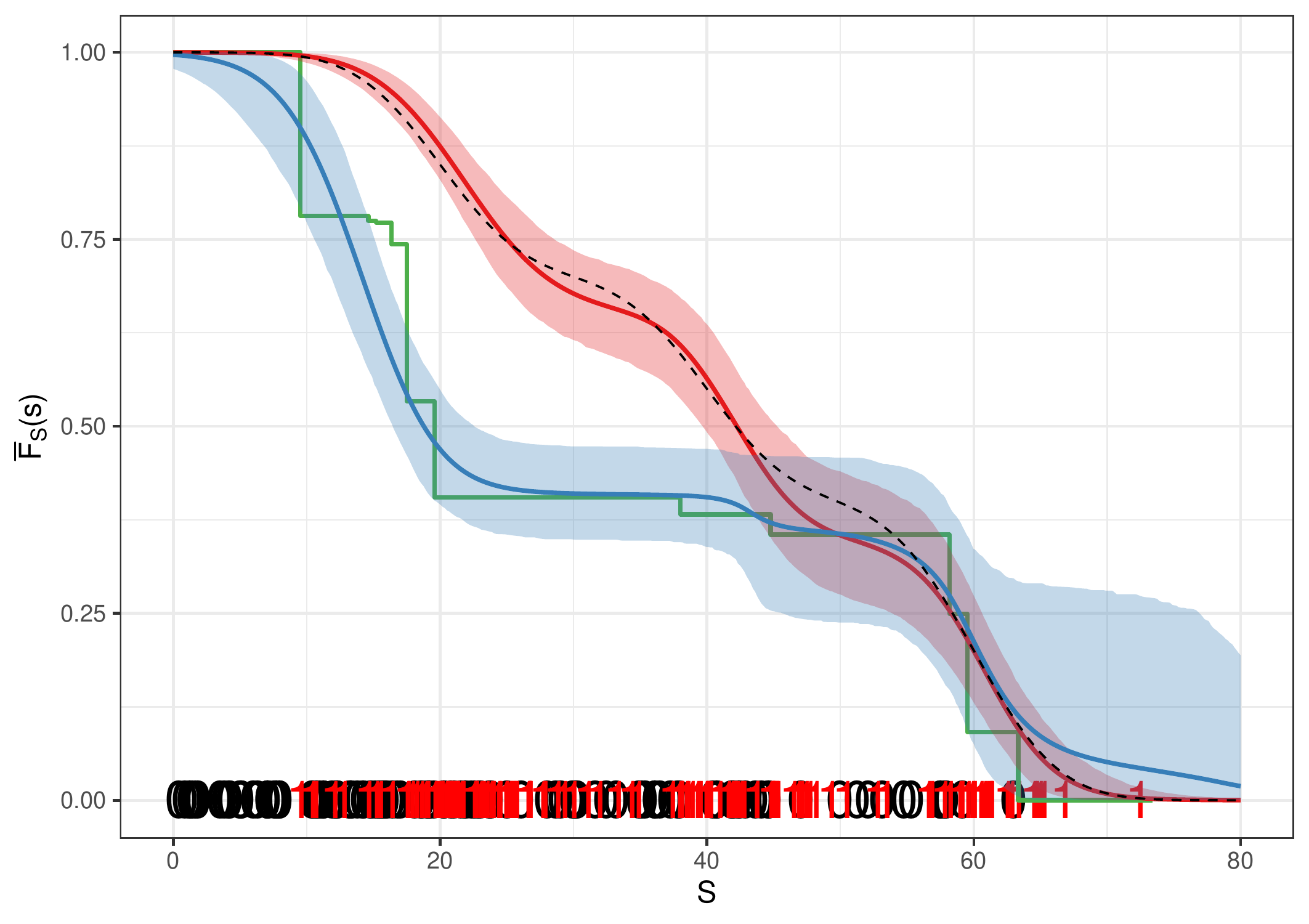}
	\caption{Simulated data: Right and left censoring times are represented by ``0'' and ``1'', respectively, on the $x$-axis. 
	The green step function shows an estimate of the survival function under the nonparametric MLE. 
	The blue curve shows an estimate of the survival function under a mixture of normals model (the simulation truth).
	The red curve shows an estimate of the survival function under the proposed model.
	Shaded areas represent pointwise $95\%$ credible intervals for the estimated survival functions.
	The black dashed line represents the simulation truth.}
	\label{fig:sim_example}
\end{figure}
Inference under the BNP-CS model for the same data used in the illustration of Section \ref{sec:EM} recovers the underlying truth better than inference under the model of the simulation truth.
Figure \ref{fig:sim_example} shows the survival function estimated under (i) an unconstrained nonparametric model estimated by the EM algorithm, (ii) a mixture of $K = 3$ normal distributions, and (iii) the proposed nonparametric model with dependent censoring. 
Although the model under (ii) matches the simulation truth, both, (i) and (ii) fail to recover a meaningful estimate, while inference under (iii) successfully exploits the information that is contained in the observed $C_{i}$.

\section{Bivariate Survival Regression for Partially Ordered Current Status Data}
\label{sec:bivariate}

\subsection{A Bivariate Event Time Model}

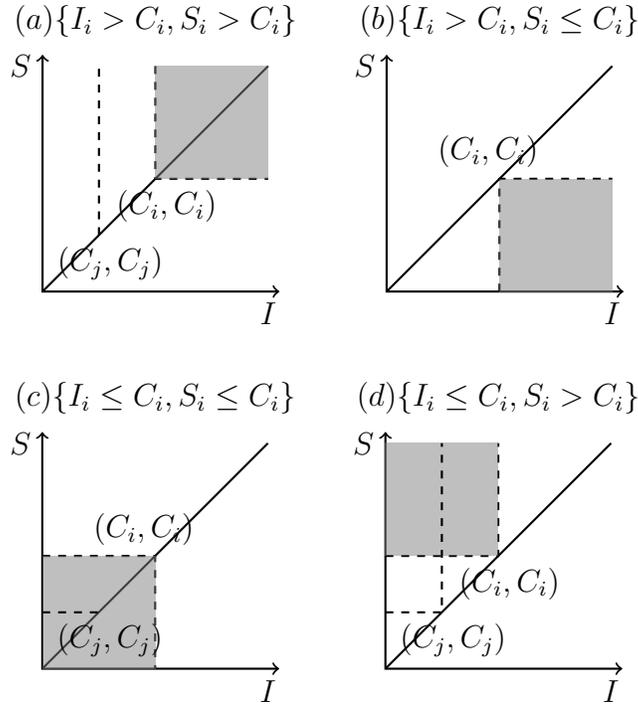
\begin{figure}[!ht]
	\begin{center}
	\begin{tikzpicture}[scale = 3]
		\draw[pattern=horizontal lines, pattern color=indianred3, draw opacity = 0] (0.25,0.25) -- (1,1) -- (0.25,1) -- (0.25, 0.25);
		\draw [thick] (1,0) node[below]{$I$};		
		\draw [thick] (0,1) node[left]{$S$};
		\draw [thick] (0.5,1.2) node{$(a) \{I_{i} > C_{i}, S_{i} > C_{i}\}$};
		\draw [<->, thick] (0,1.05) -- (0,0) -- (1.05,0);
		\draw[thick, dashed] (0.5,1) -- (0.5,0.5) -- (1,0.5);
		\draw[thick, dashed] (0.25,0.25) -- (0.25,1);
		\draw[thick] (0,0) -- (1,1);
		\draw[fill=gray, fill opacity=0.5, draw opacity = 0] (0.5,1) -- (0.5,0.5) -- (1,0.5) -- (1, 1) -- (0.5,1);
		\draw [thick] (0.55,0.5) node[below]{$(C_i,C_i)$};
		\draw [thick] (0.3,0.25) node[below]{$(C_j,C_j)$};
	\end{tikzpicture}
	\hskip10pt
	\begin{tikzpicture}[scale = 3]
		\draw [thick] (1,0) node[below]{$I$};		
		\draw [thick] (0,1) node[left]{$S$};
		\draw [thick] (0.5,1.2) node{$(b) \{I_{i} > C_{i}, S_{i} \leq C_{i}\}$};
		\draw [<->, thick] (0,1.05) -- (0,0) -- (1.05,0);
		\draw[thick, dashed] (0.5,0) -- (0.5,0.5) -- (1,0.5);
		\draw[thick] (0,0) -- (1,1);
		\draw[fill=gray, fill opacity=0.5, draw opacity = 0] (0.5,0) -- (0.5,0.5) -- (1,0.5) -- (1, 0) -- (0.5,0);
		\draw [thick] (0.45,0.5) node[above]{$(C_i,C_i)$};
	\end{tikzpicture}	
	\end{center}
	\begin{center}
	\begin{tikzpicture}[scale = 3]
		\draw [thick] (1,0) node[below]{$I$};		
		\draw [thick] (0,1) node[left]{$S$};
		\draw [thick] (0.5,1.2) node{$(c) \{I_{i} \leq C_{i}, S_{i} \leq C_{i}\}$};
		\draw [<->, thick] (0,1.05) -- (0,0) -- (1.05,0);
		\draw[thick, dashed] (0,0.5) -- (0.5,0.5) -- (0.5,0);
		\draw[thick] (0,0) -- (1,1);
		\draw[fill=gray, fill opacity=0.5, draw opacity = 0] (0,0.5) -- (0.5,0.5) -- (0.5,0) -- (0, 0) -- (0,0.5);
		\draw [above] (0.45,0.5) node[above]{$(C_i,C_i)$};
		\draw[pattern=horizontal lines, pattern color=indianred3, draw opacity = 0] (0,0.25) -- (0.25,0.25) -- (0,0) -- (0,0.25);
		\draw [thick] (0.3,0.25) node[below]{$(C_j,C_j)$};
		\draw[thick, dashed] (0,0.25) -- (0.25,0.25);
	\end{tikzpicture}
	\hskip10pt
	\begin{tikzpicture}[scale = 3]
		\draw [thick] (1,0) node[below]{$I$};		
		\draw [thick] (0,1) node[left]{$S$};	
		\draw [thick] (0.5,1.2) node{$(d) \{I_{i} \leq C_{i}, S_{i} > C_{i}\}$};
		\draw [<->, thick] (0,1.05) -- (0,0) -- (1.05,0);
		\draw[thick, dashed] (0,0.5) -- (0.5,0.5) -- (0.5,1);
		\draw[thick] (0,0) -- (1,1);
		\draw[fill=gray, fill opacity=0.5, draw opacity = 0] (0,0.5) -- (0.5,0.5) -- (0.5,1) -- (0, 1) -- (0,0.5);
		\draw [thick] (0.55,0.5) node[below]{$(C_i,C_i)$};
		\draw[pattern=horizontal lines, pattern color=indianred3, draw opacity = 0] (0,0.25) -- (0.25,0.25) -- (0.25,1) -- (0, 1) -- (0,0.25);
		\draw [thick] (0.3,0.25) node[below]{$(C_j,C_j)$};
		\draw[thick, dashed] (0,0.25) -- (0.25,0.25) -- (0.25,1);
	\end{tikzpicture}
	\end{center}
	\caption{Domain of the data $(I > 0, S > 0)$ and support for the latent times corresponding to the four cases. The gray quadrants represent the support for the latent times corresponding to the observed censoring times $(C_i, C_i)$ under $f_{IS}^\star(I, S)$. The area with red horizontal lines represents the support for the latent times under $f_{IS}^{\prime}(I, S)$.}
	\label{fig:domain}
\end{figure}

We now use the BNP-CS model \eqref{eq:model_uni} as a building block for bivariate outcomes. 
Beyond the already discussed dependence of $S_{i}$ and $C_{i}$, we add some more structure based on prior knowledge of the underlying process. 
We introduce a mixture model to reflect that patients can experience the symptoms due to the disease of interest or they can arise from other causes, in which case we assume independence in the corresponding submodel.
That is, we model the bivariate event time distribution $f_{IS}(I, S)$ of time to infection and time to symptoms as a mixture model in which one of the two components is subject to the order constraint $I < S$. 
This translates to 
\begin{equation}
    \begin{split}
        f_{IS}(I,S) &= w f_{IS}^{\star}(I,S) + (1 - w) f^{\prime}_{IS}(I,S) 
    \end{split}
    \label{eq:joint_model}
\end{equation}
where $f^{\prime}_{IS}(I,S)$ is subject to $I < S$, whereas $f_{IS}^{\star}(I,S)$ is not. 
Therefore, $f^{\star}_{IS}(I,S)$ can be interpreted as the distribution of $(I, S)$ for a patient with symptoms ``due to other causes''.
Figure \ref{fig:domain} shows the support of the two components of the mixture as well as the support for the latent times corresponding to the four possible censoring indicators.

We make two main assumptions: (i) under $f^{\star}_{IS}(I,S)$, the time to symptoms (due to other causes) and time to infection are independent; (ii) under $f^{\prime}_{IS}(I,S)$, the latency time $L = S - I$ and the time to infection are independent.
Here $L$ is the delay from the onset of illness to the development of symptoms.
The assumed marginal $f_{I}(\cdot)$ on $I$ is shared by both, $f^{\star}_{IS}$ and $f^{\prime}_{IS}$.
Thus, model \eqref{eq:joint_model} becomes
\begin{equation}
    f_{IS}(I,S) = w f_{I}(I) f_{S}^{\star}(S) + (1 - w) f_{I}(I) f_{L}(S - I).
    \label{eq:joint_model_2}
\end{equation}
For later reference we note that sampling, $(I_{i}, S_{i}) \sim f_{IS}$ can be equivalently written as a hierarchical model with latent indicators, say $r_{i}^{(W)}$ with $p(r_{i}^{(W)} = 1) = w$ and
\begin{equation}
    (I_{i}, S_{i} \mid r_{i}^{(W)}) \sim
    \begin{cases}
        f_{I}(I) f_{S}^{\star} (S) &\text{if } r_{i}^{(W)} = 1
        \\
        f_{I}(I) f_{L} (S - I) &\text{if } r_{i}^{(W)} = 0.
    \end{cases}
    \label{eq:hier_model}
\end{equation}

The second component in \eqref{eq:joint_model_2} includes the constraint $I < S$ as a positivity constraint on the latency time $L > 0$.
Recent approaches to deal with hard constraints use relaxation methods that replace the hard constraint with priors that penalize departures outside of the constraint subspace \citep{duan2018bayesian}. 
Alternatively, \citet{patra2018constrained} developed methodology that uses unconstrained inference and then projects the posterior draws onto the constrained subspace.
In our model, assigning positive support to the reparametrized variable $L$ automatically ensures the required order constraint $I < S$.
In the following, we will use $L \mid \lambda_{L} \sim \Exp (\lambda_{L})$.
As a consequence, under the component $f^{\prime}_{IS}(I,S) = f_{I}(I) f_{L}(S - I)$, time to symptoms and time to infection are dependent.

We highlight our use of structural model assumptions that reflect prior knowledge of the biology behind current status data for infectious diseases.
In particular, as discussed thoroughly in \citet{wang2000assessing}, the joint distribution for bivariate current status data is not likelihood identifiable nonparametrically.
Thus, two approaches are possible: (i) estimate the joint distribution under parametric or semiparametric assumptions, or (ii) build the joint model from the two identifiable marginal distributions and a particular choice for their dependence structure.
This article uses the latter strategy.

\subsection{Bayesian Nonparametric Priors}

The model is completed by introducing priors for the two unknown distributions, assuming nonparametric mixture models for both $f_{I}(I)$ and $f_{S}^{\star}(S)$,
\begin{equation}
    \begin{split}
        &f_{I}(I) = \int \text{N}(I \mid \btheta^{(I)}) dH^{(I)}(\btheta^{(I)}) = \sum_{k=1}^{+\infty} \pi_{k}^{(I)} \text{N} (I \mid \mu_{k}^{(I)}, \sigma_{k}^{(I)2})
        \\
        &f_{S}^{\star}(S) = \int \text{N}(I \mid \btheta^{(S)}) dH^{(S)}(\btheta^{(S)}) = \sum_{k=1}^{+\infty} \pi_{k}^{(S)} \text{N} (S \mid \mu_{k}^{(S)}, \sigma_{k}^{(S)2}),
    \end{split}
    \label{eq:marginal_models}
\end{equation}
where $\btheta^{(I)} = (\mu^{(I)}, \sigma^{(I)2})$ and $\btheta^{(S)} = (\mu^{(S)}, \sigma^{(S)2})$.
Here $H^{(I)}(\cdot) = \sum_{k} \pi_{k}^{(I)} \delta_{\btheta^{(I)}}$, and similarly $H^{(S)}$, are the random mixing measures. 
The model is completed with a prior probability model on $H^{(I)}$ and $H^{(S)}$. 
Prior distributions on random probability measures are known as nonparametric Bayes (BNP) models.

Using a nonparametric prior on $H^{(I)}$ and $H^{(S)}$ the model becomes a mixture of normals with respect to the chosen random mixing measure. 
For example, in our implementation we assume a DP prior again, as in \eqref{eq:model_uni}, now using two instances for $f_{I}$ and $f_{S}^{\star}$. 
Alternatively, any other nonparametric Bayesian prior \citep[e.g.][]{james2009posterior} could be used. 
The following result gives the marginal distributions implied by our construction.
\begin{Thm}
The marginal distributions implied by model \eqref{eq:joint_model_2} with priors \eqref{eq:marginal_models} are
\begin{align*}
&f_{I}(I) = \sum_{k=1}^{+\infty} \pi_{k}^{(I)} \mathrm{N} (I \mid \mu_{k}^{(I)}, \sigma_{k}^{(I)2})
\\
&f_{S}(S) = w \sum_{k=1}^{+\infty} \pi_{k}^{(S)} \mathrm{N} (S \mid \mu_{k}^{(S)}, \sigma_{k}^{(S)2}) + (1 - w) \sum_{k=1}^{+\infty} \pi_{k}^{(I)} \mathrm{EMG} (S \mid \mu_{k}^{(I)}, \sigma_{k}^{(I)2}, \lambda_{L}),
\end{align*}
where $\mathrm{EMG}(\mu, \sigma^{2}, \lambda)$ denotes the exponentially modified Gaussian distribution \citep{grushka1972characterization}. 
\end{Thm}
Model \eqref{eq:joint_model_2} together with \eqref{eq:marginal_models} and \eqref{eq:model_uni} for $p(C_{i} \mid S_{i})$ defines the proposed \textbf{bivariate BNP-CS model} for current status data.

One of the reasons for the wide use of BNP mixtures like \eqref{eq:marginal_models} is the induced prior on a random partition.
Consider $I_{i} \sim f_{I}$, $i = 1, \dots, n$. Under model \eqref{eq:marginal_models} we can introduce latent indicators, say $r^{(I)}_{i}$, and write instead
\begin{equation*}
    p(I_{i} \mid r^{(I)}_{i} = k) = \mathrm{N}(\mu^{(I)}_{k}, \sigma^{(I)2}_{k}) \quad \text{and} \quad p(r^{(I)}_{i} = k) = \pi_{k}^{(I)}.
\end{equation*}

The $r^{(I)}_{i}$'s can be interpreted as cluster membership indicators. 
We see then how this formulation implicitly defines a probability model $p(\br^{(I)})$ on a partition $\br^{(I)} = (r^{(I)}_{1},\dots,r^{(I)}_{n})$.
Two observations are clustered together if they are assigned the same group-specific parameters $\btheta_{k} = (\mu_{k}, \sigma^{2}_{k})$, where for brevity we now omit the superscript $(I)$.
Recall the indicators $r_{i}^{(W)}$ in \eqref{eq:hier_model}. 
Without loss of generality assume that $r_{i}^{(W)} = 1$ (symptoms due to other causes) for $i = 1,\dots, n_{1}$, and $r_{i}^{(W)} = 0$ (symptoms due to disease), $i = n_{1} + 1,\dots, n$. 
Similar to $p(\br^{(I)})$ we get a random partition $p(\br^{(S)}_{\star})$ induced by sampling from $f_{S}^{\star}(\cdot)$ for patients $i = 1,\dots,n_{1}$.
Analogously, for patients $i = n_{1} + 1, \dots, n$ we have a partition $\br_{S} =(r_{i}^{(S)},i = n_{1} + 1,\dots,n)$, with $r_{i}^{(S)} =r_{i}^{(I)}$ due to $S_{i} = I_{i} + L_{i}$.
In words, under the proposed model, the clustering structures $\br^{(S)}$ and $\br^{(S)}_{\star}$ for symptoms due to infection and for symptoms due to other causes, respectively, are modeled separately and are independent. 
In fact, symptoms due to infection inherit the clustering structure $\br^{(I)}$, which is induced by the marginal distribution for the infection times.

In order to cluster grouped data, other approaches have been proposed \citep{teh2005sharing, rodriguez2008nested, camerlenghi2019distribution, argiento2020hierarchical}.
These strategies allow for the possibility of sharing atoms of the random probability measures across groups, thus borrowing information and yielding more precise inference. 
However, the random partition is not the main inference target here and we shall therefore not further explore such alternatives.

\subsection{Regression on Covariates}

We now add covariate effects in the proposed nonparametric model. 
In the context of model \eqref{eq:marginal_models} this takes the form of replacing $H^{(I)}$ and $H^{(S)}$ by families of random probability measures (r.p.m). 
That is, we introduce a family $\{H^{(I)}_{\bx}, \bx \in \mathcal{X} \}$, and similarly for $H^{(S)}$.
Here $\bx$ are patient specific covariates, and we replace $H^{(I)}$ and $H^{(S)}$ by $H^{(I)}_{\bx_{i}}$ and $H^{(S)}_{\bx_{i}}$ for patient $i$ in equation \eqref{eq:marginal_models}.
Dropping for the moment the superscript for easier exposition, let $\mathcal{H} = \{H_{\bx} = \sum_{k} \pi_{xk} \delta_{\mu_{xk}}, \bx \in \mathcal{X}\}$ denote a family of r.p.m.'s indexed by $\bx$.
The most widely used class of priors on families like $\mathcal{H}$ are dependent DP (DDP) models \citep{maceachern1999dependent}.
A recent review appears in \citet{quintana2020dependent}.
The DDP construction implies marginally for each $H_{\bx}$ a DP prior, and allows for the desired dependence across $\bx$. 
The definition of the marginal DP implies that the $\mu_{xk}$'s are independent across $k$ and that the weights have stick-breaking priors, but it does not restrict the distribution across $\bx$.
This is what the DDP construction exploits to borrow information across covariate values.
The DDP induces dependence across $\bx$ through the atoms $\mu_{xk}$ and/or the weights $\pi_{xk}$ of the marginal r.p.m.'s.
In \citet{maceachern1999dependent}, dependence is induced by assuming that, for fixed $k$, the atoms $\mu_{xk}$ are realizations of a Gaussian process, indexed by $\bx$. 
Independence across $k$, together with the stick-breaking prior for the common weights $\pi_{k}$ (not indexed by $\bx$), maintains the marginal DP prior on $H_{\bx}$.
This instance of the DDP model is known as ``common weights DDP.''
Alternative implementations are possible, with dependent (across $\bx$) weights $\pi_{xk}$ and common atoms (``common atoms DDP''), or the most general DDP model with dependent weights and atoms.

In the Partner Notification study the predictors are $\bx_{i} = \{\text{gender}, \text{arm}, \text{age}\} \in \{0; 1\}^2 \times \mathbb{R}^{+}$, i.e. two binary and one continuous covariate. 
We use a simple ANOVA structure to induce dependence of $\mu_{xk}$ across $\bx$. 
DDP models with ANOVA-type dependence across categorical factors are introduced as the ANOVA-DDP in \citet{de2004anova} and then extended to continuous covariates in \citet{de2009bayesian}.
The dependence structure of the random probability measures $H_{\bx}$ is modeled by constructing the atoms as $\mu_{xk} = \delta_{k} + \alpha_{k} x_{1} + \beta_{k} x_{2} + \gamma_{k} x_{3}$.
The interpretation of the linear model coefficients $\bm_{k} = (\delta_{k}, \alpha_{k}, \beta_{k}, \gamma_{k})^{\intercal}$ is exactly as in an ANOVA model, inducing the desired dependence of $H_{\bx}$ across $\bx$ by sharing, for example, the same $\beta_{k}$ for any two covariate vectors $\bx$ and $\bx^{\prime}$ that share the same $x_{2}$. 
Finally, using a design vector $\bd_{i} = (1, x_{i1}, x_{i2}, x_{i3})^{\intercal}$ to select the desired ANOVA effects we can write $\mu_{x_{i}k} = \bd_{i}^{\intercal} \bm_{k}$ to get $H_{\bx_{i}} = \sum_{k=1}^{+\infty} \pi_{k} \delta_{\bd^{\intercal} \bm_{k}}$.
Defining $\btheta_{k} = (\bm_{k}, \sigma_{k}^{2})^{\intercal}$ to allow for a mixture also with respect to the kernel variances, and defining one common mixing measure
    $$H(\cdot) = \sum_{k=1}^{+\infty} \pi_{k} \delta_{\btheta_{k}}$$
the marginal distribution $f_{I}(I_{i} \mid \bx_{i})$ can be rewritten equivalently as a DP mixture of linear models, now using a single mixing measure $H$ for all $\bx$ \citep[linear dependent DDP, ][]{jara2010bayesian}.
Also, we add back the superscripts $(I)$ and $(S)$ on the model parameters:
\begin{equation}
	f_{I} (I_i \mid \bx_{i}) = \int \text{N}(I_i | \bd_{i}^{\intercal} \bm^{(I)}, \sigma^{(I)2}) dH^{(I)} (\btheta^{(I)}) \quad \mathrm{with} \quad H^{(I)} \sim \text{DP}(M^{(I)}, H_0^{(I)}).
	\label{eq:final_model}
\end{equation}
Another instance of the same model is used for the marginal distribution of symptoms due to other causes $f_{S}^{\star}(S_{i} \mid \bx_{i})$.
The full model is 
\begin{equation*}
    \begin{split}
	    &C_{i} \mid S_i, \lambda = \min \{ S_{i} + \Exp(\lambda); \Unif(A, B) \}
	    \\
	    &(S_{i}, I_{i}) \mid \btheta^{(S)}, \btheta^{(I)}, w, \lambda_{L} \sim f_{IS}(I,S).
    \end{split}
\end{equation*}
using \eqref{eq:final_model} for $f_{I}$ and similarly for $f_{S}^{\star}$.
The complete model now defines a \textbf{bivariate BNP-CS survival regression}.
Using the stick-breaking representation, the DP priors on $H^{(I)}$ and $H^{(S)}$ can be written as follows. 
Using superscripts $E \in \{I,S\}$ to refer to the construction of $f_{I}$ and $f_{S}^{\star}$ respectively, we have
\begin{equation*}
    \begin{split}
        &\{ \bm_{k}^{(E)}, \sigma_{k}^{(E)2}\}_{k = 1}^{+\infty} \mytilde{iid} H_0^{(E)} =  \text{N}(\bm_{k}^{(E)} \mid \bm_0^{(E)}, \Sigma_{0}^{(E)}) \times \text{IG}(\sigma_{k}^{(E)2} \mid a_\sigma^{(E)}, b_\sigma^{(E)})
	    \\
	    &\bpi^{(E)} \mid M^{(E)} \sim \text{SB}(M^{(E)}); \quad M^{(E)} \sim \text{Ga}(a_{M}, b_{M}),
    \end{split}
\end{equation*}
and $\lambda \sim \text{Ga}(a_{\lambda}, b_{\lambda})$, $\lambda_{L} \sim \text{Ga} (a_{L}, b_{L})$, $w \sim \Beta (a_{w}, b_{w})$.

For later reference we note that the random probability measures $H^{(I)}(\btheta^{(I)})$ and $H^{(S)}(\btheta^{(S)})$ that serve as the mixing measure in \eqref{eq:final_model} are multivariate distributions for $\btheta^{(I)} = (\bm^{(I)}, \sigma^{(I) 2})^{\intercal} = (\delta^{(I)}, \alpha^{(I)}, \beta^{(I)}, \gamma^{(I)}, \sigma^{(I) 2})^{\intercal}$, and similarly for $\btheta^{(S)}$.
Let 
\begin{equation}
    H^{(I)}_{\beta} = \sum_{k=1}^{+\infty} \pi_{k}^{(I)} \delta_{\beta_{k}^{(I)}} 
    \label{eq:marg_eff}
\end{equation}
denote the implied univariate marginal for the ANOVA effect $\beta^{(I)}$.
Analogous notation can be used for $H^{(S)}_{\beta}$ and any of the other ANOVA effects. 
We will later use inference on $H^{(E)}_{\beta}$, $E \in \{I,S\}$, to summarize inference on the treatment effect.

\section{Posterior Inference}
\label{sec:post_inference}

To implement posterior inference under a Dirichlet process mixture model, the two main strategies are marginal \citep{escobar1995bayesian, maceachern1998estimating, neal2000markov} and conditional \citep{ishwaran2001gibbs, kalli2011slice} MCMC posterior simulation. 
In our implementation, we employ the latter.
In particular, we rewrite the mixture model as a hierarchy by explicitly introducing the latent cluster membership variables $\br^{(W)}$, $\br^{(I)}$ and $\br^{(S)}_{\star}$.
Moreover, we impute the latent symptoms and infection times from their corresponding full conditionals. 
We use efficient sampling for truncated normal distributions, originally proposed in \citet{geweke1991efficient}.
This allows us to use standard algorithms for inference under a DPM.

The total masses for the two random probability measures, $M^{(I)}$ and $M^{(S)}$, are included in the MCMC scheme and assigned Gamma priors, as recommended in \citet{escobar1995bayesian}.
Moreover, we put priors on the hyperparameters for the base measures $H_{0}^{(I)}$ and $H_{0}^{(S)}$.
Additional details of the algorithm are deferred to Section \ref{sec:mcmc} in the supplementary materials.

\section{Partner Notification Study - Results}
\label{sec:application}

We apply the proposed model for inference in the Partner Notification study described in Section \ref{sec:real_Study}.
The primary inference goal is to understand the effect of covariates, in particular treatment assignment, on the joint distribution of the two latent times of interest.
Furthermore, we are interested in assessing what factors drive time to rehospitalization with infection and how time to symptoms onset of these cases can improve such estimation.

\begin{figure}[!ht]
	\centering
	\includegraphics[width=.49\linewidth]{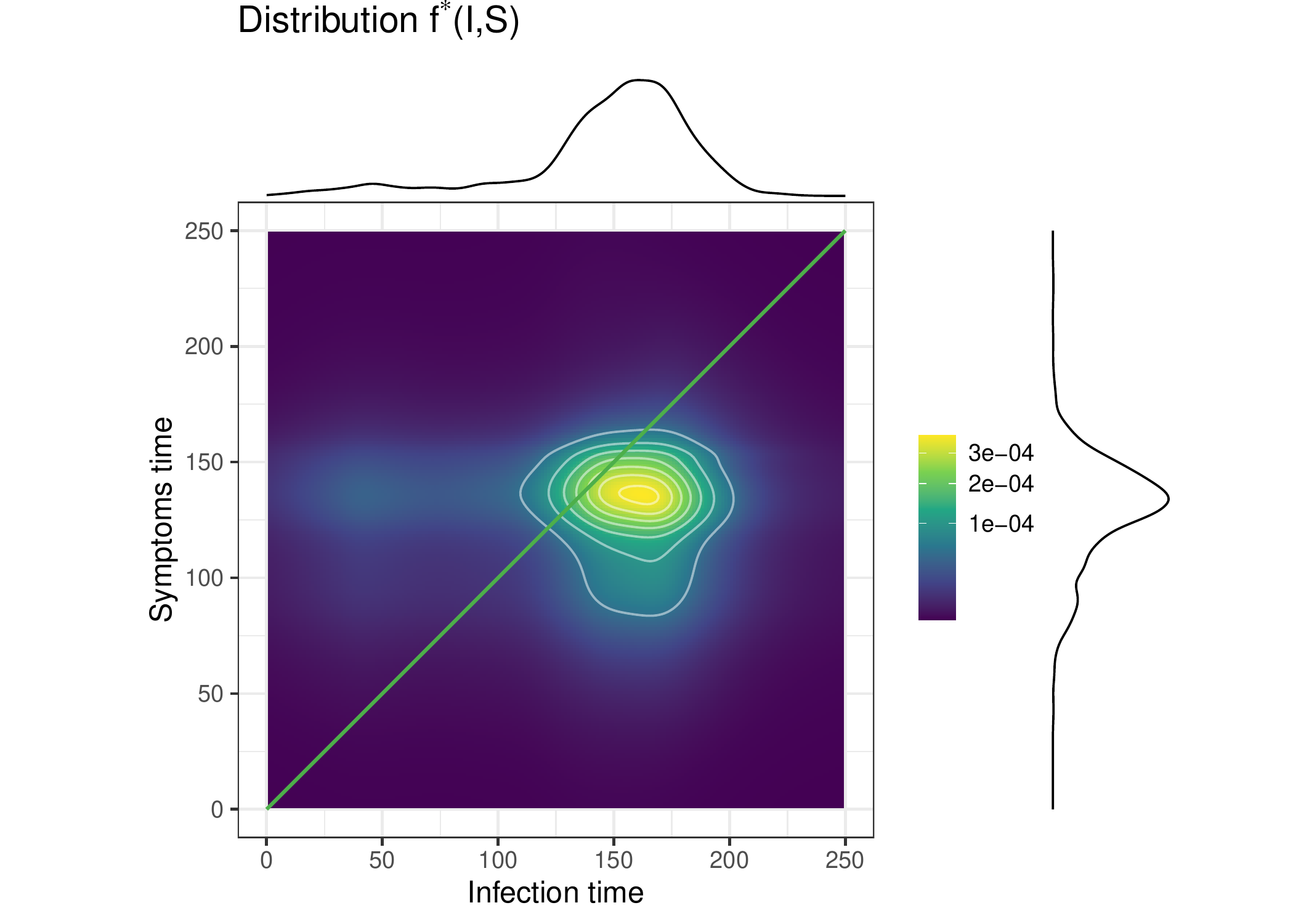}
	\includegraphics[width=.49\linewidth]{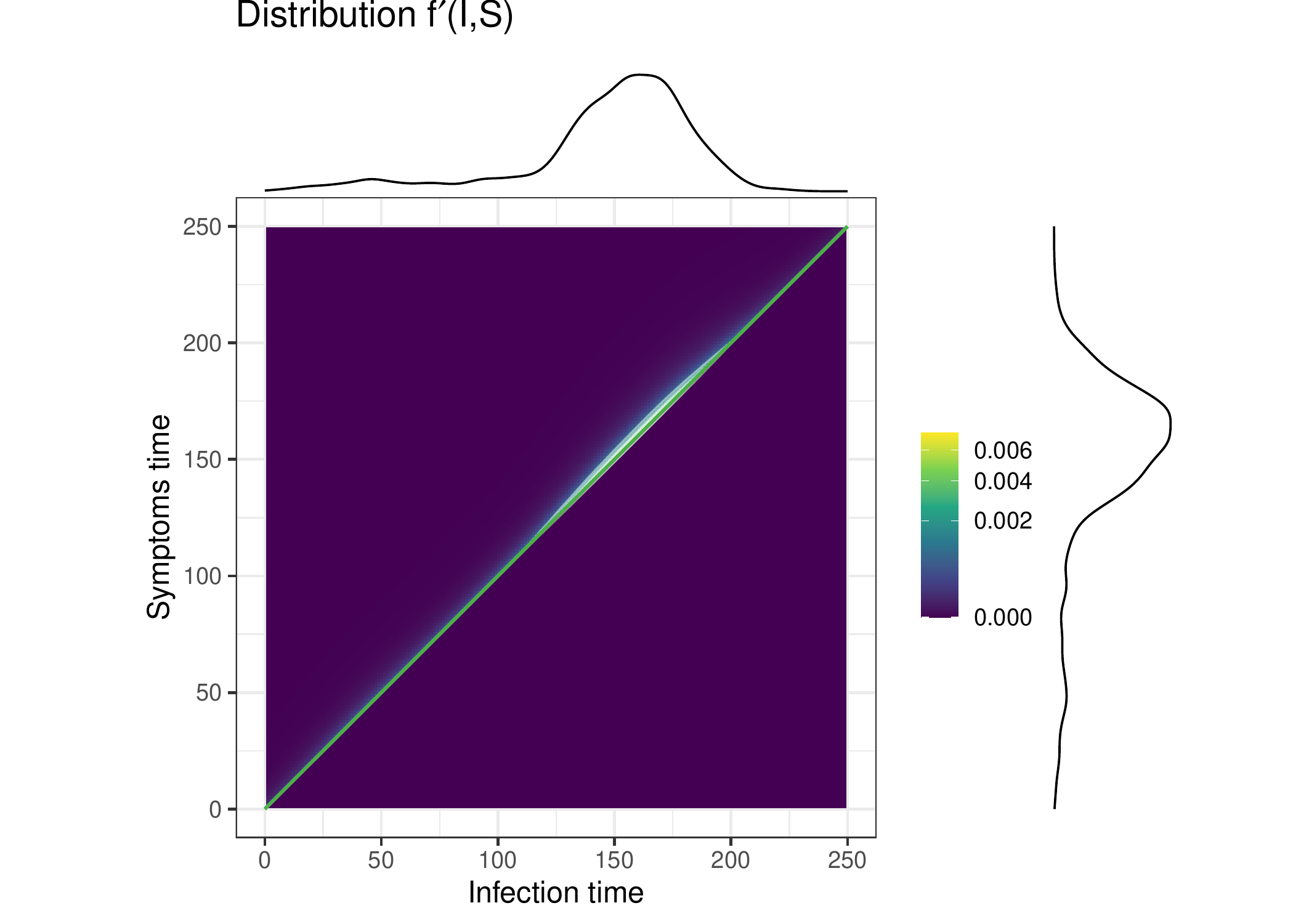}
	\includegraphics[width=.49\linewidth]{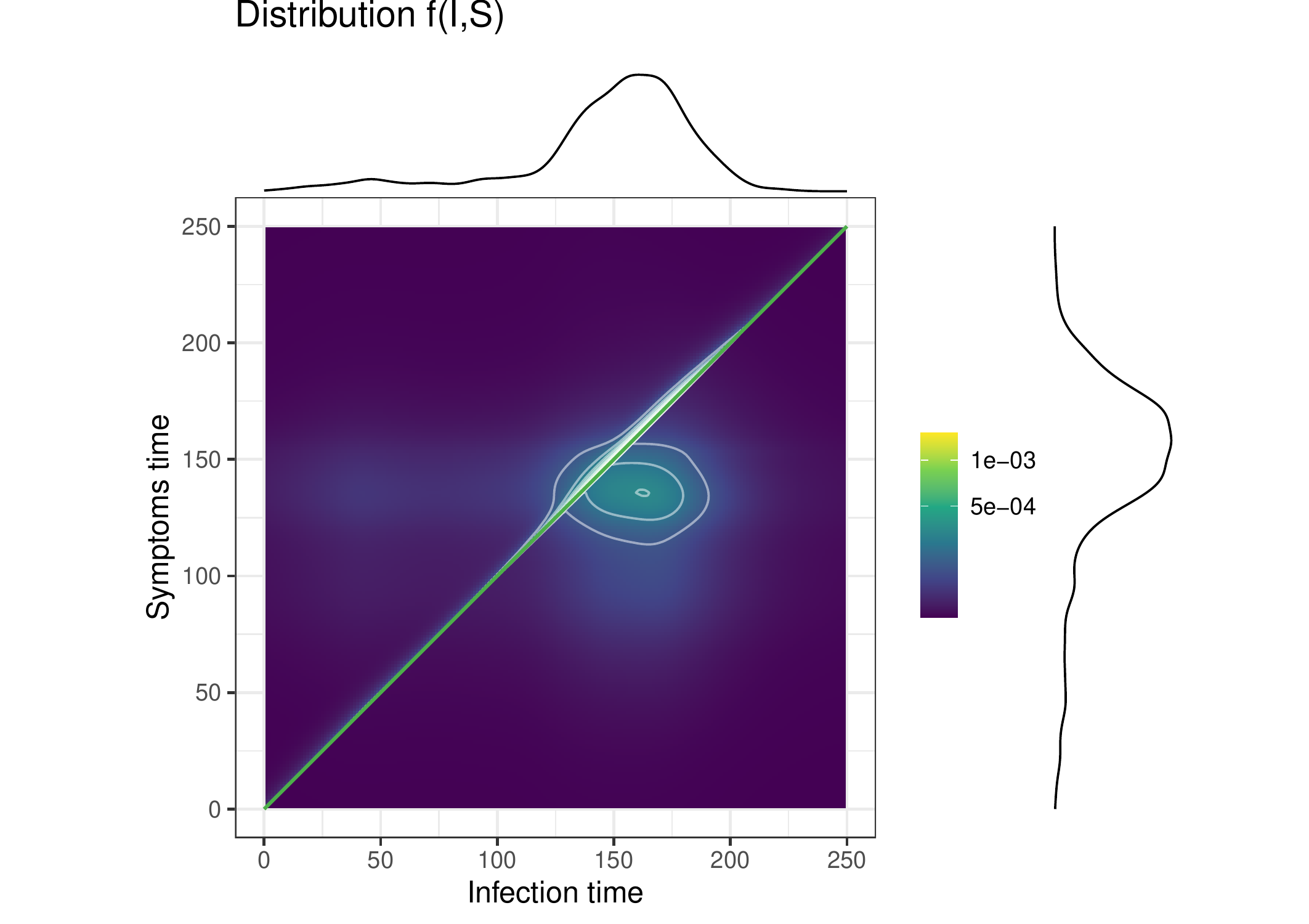}
	\caption{Results: Posterior mean density estimate for $f_{IS}^{\star}$, $f_{IS}^{\prime}$ and $f_{IS}$ corresponding to the baseline covariate levels (male, control group, mean age). The green line corresponds to the $45^{\circ}$ line, i.e. $I = S$.
	The corresponding marginal distributions are shown on the top and right side of the density plot.}
	\label{fig:real_density}
\end{figure}

Inference under the proposed model includes the full joint distribution of latent times to symptoms and infection times.
Figure \ref{fig:real_density} shows the posterior estimated distribution $f_{IS}(I,S)$ and the two components $f_{IS}^{\star}(I,S)$, $f_{IS}^{\prime}(I,S)$ corresponding to a `baseline' covariate combination (male, control group, median age). 
There is significant probability mass in the lower triangle ($S < I$) that is not concentrated around the $45^{\circ}$ line but is quite spread out. 
Instead, for the constrained component ($S > I$) the probability mass is concentrated very close to the $45^{\circ}$ line. 
In other words, most of the inferred symptoms times due to infection concentrate in $I < S < I + 10$. 
This is coherent with the fact that we expect the symptoms due to the infection to follow shortly after the disease onset.

\begin{figure}[!ht]
	\centering
	\includegraphics[width=.75\linewidth]{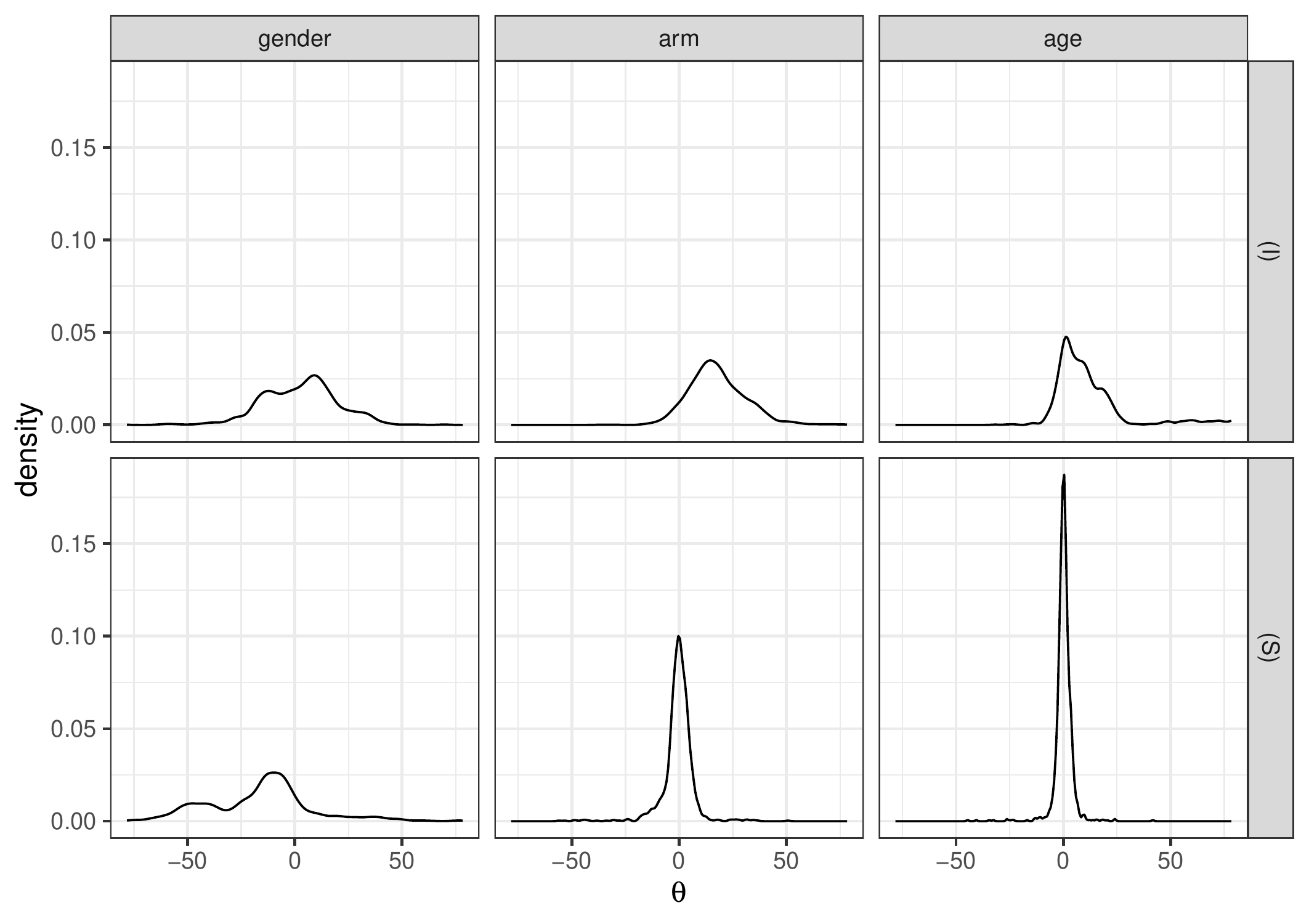}
	\caption{Results: Estimated distributions $H_{\alpha}, H_{\beta}$ and $H_{\gamma}$ for the regression coefficients $\alpha$ (left), $\beta$ (middle) and $\gamma$ (right) under $f_{I}$ (top panels) and $f_{S}^{\star}$ (bottom panels).}
	\label{fig:cov_effects}
\end{figure}

To show the estimated covariate effects, we could compare density estimates for different combinations of the predictors.
Alternatively, we can report posterior estimates of the marginal distributions for the ANOVA effects, for example $H^{(E)}_{\beta}$, $E \in \{I, S\}$ from \eqref{eq:marg_eff}. 
These are the univariate marginal distributions of the treatment effect in the DDP model, and concisely summarize the change of bivariate survival distribution with respect to treatment versus control.
The top center panel in Figure \ref{fig:cov_effects} shows the posterior estimated distributions $\mathbb{E}(H^{(I)}_{\beta} \mid \text{data})$, and similarly for other regression effects.
Two significant effects can be detected.
Importantly, the treatment group seems to have delayed infection times, confirming what was found in an earlier analysis in \citet{saly2011nonparametric}. 
Moreover, gender seems to have an effect on the time to symptoms due to other causes, with women seeking early hospital visits because of their symptoms.
This might be simply due to the fact that women are more aware of their symptoms and are more inclined to hospital visits, suggesting that a health education campaign for men might improve their health outcome. 
Age has also been found to have a weak effect: younger individuals have shorter infection times, possibly due to their more risky behaviour.

Two parameters of the model, namely $\lambda_{L}$ and $\lambda$, can give insights into how long it takes for participants to develop symptoms and to seek a visit to the hospital. 
In particular, the $95\%$ credible interval for the exponential parameter $\lambda$ is $[0.70, 1.42]$, suggesting that people seek re-hospitalization, on average, one day after they start developing symptoms.
Moreover, the $95\%$ credible interval for the exponential parameter $\lambda_{L}$ is $[0.22, 0.80]$, which implies that patients develop symptoms due to infection, on average, $2.5$ days after the infection onset.

Inference includes an estimate for the proportion of patients that experience symptoms due to the infection, in our notation $1 - w$. 
The posterior mean of such proportion is $17.72\%$ ($95\%$ CI: $[11.71\%$, $24.06\%$]).
This is coherent with what we see empirically in the data. 
There are more observed symptoms than observed infections, which implies that most of the symptoms should be attributed to other causes. 
This finding has important practical implications as it can help better planning for the treatment of patients.

\begin{figure}[!ht]
	\centering
	\includegraphics[width=.49\linewidth]{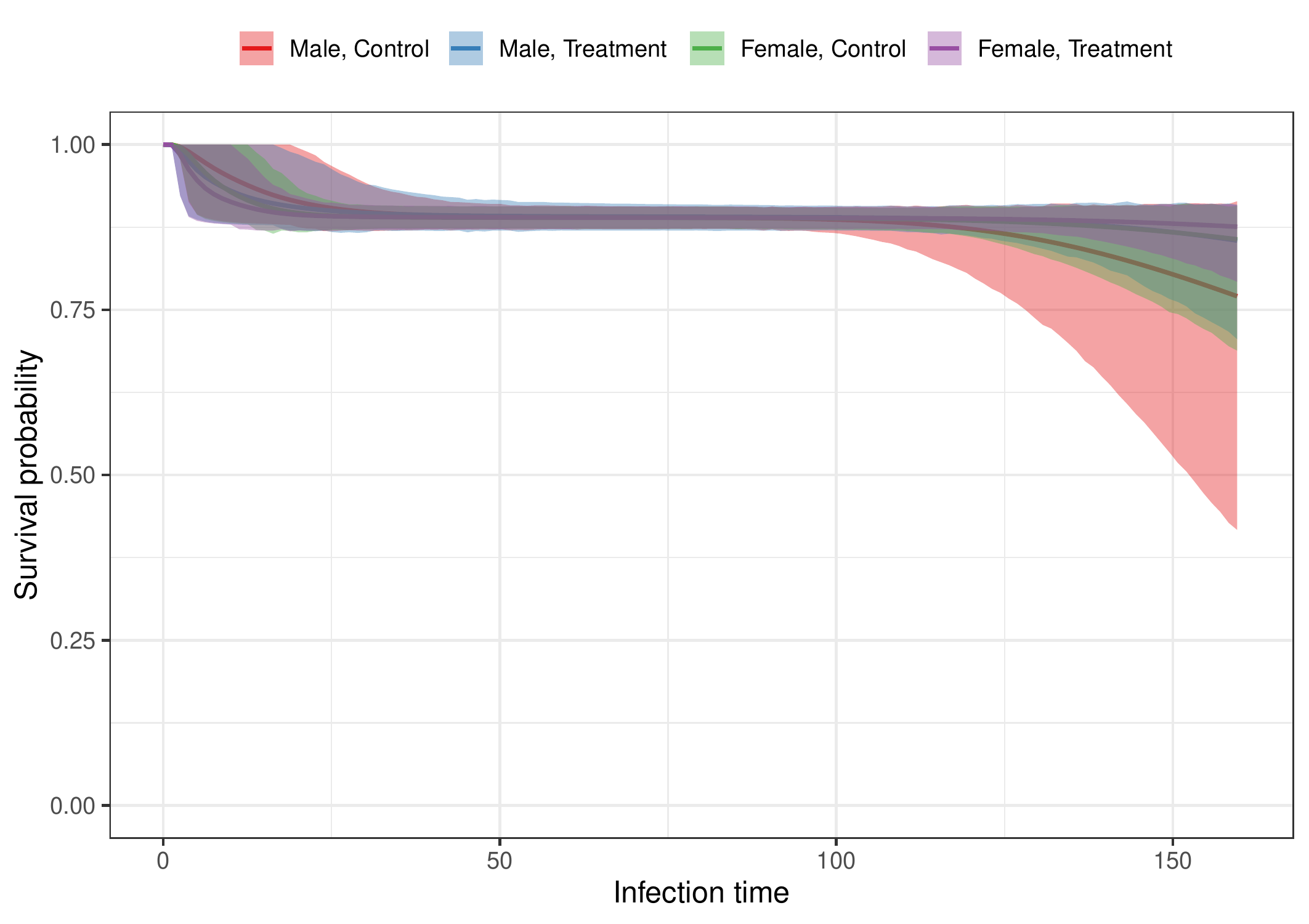}
	\includegraphics[width=.49\linewidth]{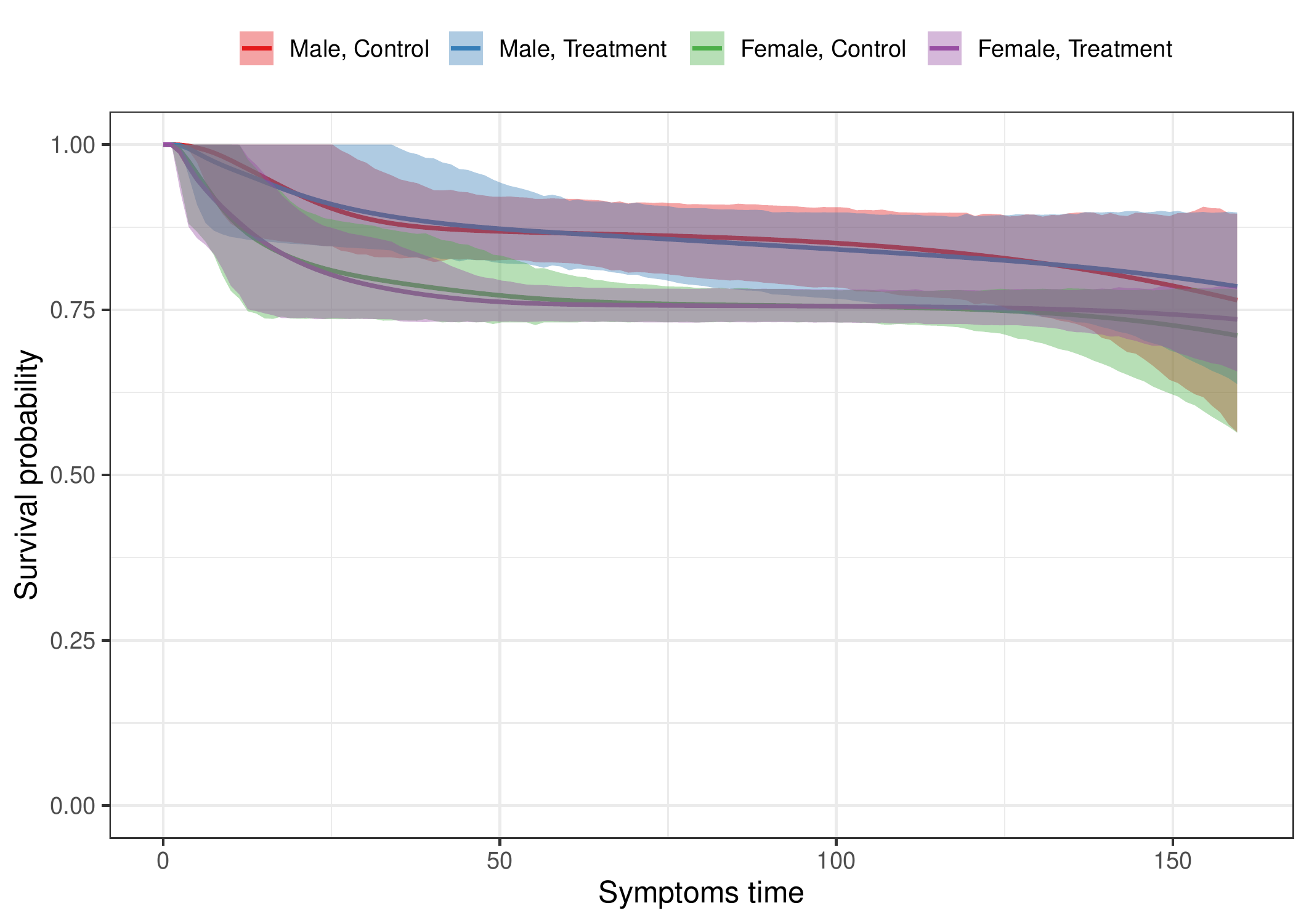}
	\caption{Results: LDDP estimated survival curves for infection times (left panel) and times until symptoms (right panel) corresponding to the possible combinations of the binary covariates \textit{gender} and \textit{treatment} fixing the predictor \textit{age} to the average age in the sample.}
	\label{fig:LDDP_joint}
\end{figure}

We compare with alternative inference under two independent linear dependent Dirichlet process (LDDP) mixture of survival models for the marginal distributions of infection and symptoms times.
This method is described in \citet{de2009bayesian} and implemented in the \texttt{DPpackage} \citep{dppackage}.
For a fair comparison, we used the same prior specifications for the shared parameters under the two models.
The results are shown in Figure \ref{fig:LDDP_joint}.
Consistent results can be found across the two models. 
For example, under the estimated models women have shorter time until symptoms as measured by the distribution for the corresponding regression coefficient in Figure \ref{fig:cov_effects} and by the survival curve in Figure \ref{fig:LDDP_joint} (right).
Unlike inference under the marginal models, inference under the proposed bivariate model also shows an effect of the treatment on the infection time.
Patients in the intervention group have a delayed re-infection time. 
The proposed model yields more interpretable results compared to the two independent LDDP models.
In fact, under the latter models the probability mass accumulates toward the bounds of the observed censoring times, yielding a ``flat'' survival curve in the middle region (see Figure \ref{fig:LDDP_joint}), exactly where we expect events to happen. 
This shows that the prior shrinkage alone does not suffice for regularization, and it is consistent with the observations of Section \ref{sec:EM}.
In fact, most right censored observations are imputed to the right of the rightmost censoring time, whereas most left censored observations are imputed to the left of the leftmost censoring time.

\section{Discussion}
\label{sec:discussion}
 
We proposed a novel Bayesian nonparametric bivariate survival regression model that is especially suited for current status data (BNP-CS regression).
This research was motivated by the failure of available methods for such data formats.
For example, we showed that widely used nonparametric mixture priors lead to biologically uninterpretable results.
Our model was built by incorporating simple structural dependence assumptions in a linear dependent Dirichlet process mixture of survival models.

Applied to a recurrent infection study, the method provides novel insights into how symptoms-related hospital visits are affected by covariates. 
Notably, we were able to replicate previous results showing a significant effect of the intervention in the randomized clinical trial under consideration.
In particular, patients in the intervention group have an improved outlook as measured by delayed reinfections. 
We also detect an effect of age, with young people having earlier reinfections, which might be due to more risky behaviours. 
Furthermore, we show that gender has a significant effect on the time until symptoms, but not on infection times.
Our study shows that men seek hospital visits later compared to women, suggesting that investing in an awareness campaign could be beneficial.

The ideas presented in this article can be extended to different dependence structures. 
The present data called for a positive correlation between infection times and infection-related symptom times. 
A similar model specification can be used for negative correlations. 
Once the marginal models are flexibly specified, one could for example use copula models to construct a joint distribution with the desired dependence structure.
A similar approach, but with positive correlations, could be used for general positively correlated event times when the assumptions used in this application are not available.

\section*{Supplementary Materials}

Supplementary materials present additional details.
These include proofs of the theorems, the MCMC scheme, convergence diagnostics and simulation studies.
In separate files, the supplementary materials additionally include the \texttt{R} programs implementing
the model developed in this article.

\section*{Acknowledgments}

Dr. M\"uller acknowledges partial support from grant NSF/DMS 1952679 from the National Science Foundation, and under R01 CA132897 from the U.S. National Cancer Institute. 
Dr. Sal y Rosas Celi was supported by Direcci\'{o}n de Gesti\'{o}n de la Investigaci\'{o}n at the PUCP through grant DGI-2017-496.

\vspace{1cm}
\baselineskip=14pt
\bibliographystyle{natbib}
\bibliography{biblio}

\begin{thebibliography}{}

\bibitem[Escobar and West(1995)Escobar and West]{escobar1995bayesian}
Escobar, M.~D. and West, M. (1995).
\newblock Bayesian density estimation and inference using mixtures.
\newblock {\em Journal of the American Statistical Association\/}, {\bf 90},
  577--588.

\bibitem[Geweke(1991)Geweke]{geweke1991evaluating}
Geweke, J. (1991).
\newblock Evaluating the accuracy of sampling-based approaches to the
  calculation of posterior moments.
\newblock In {\em Proceedings of the Fourth Valencia International Conference
  on Bayesian Statistics\/}, pages 169--193.

\end{thebibliography}


\begin{thebibliography}{}

\bibitem[Argiento {\em et~al.}(2020)Argiento, Cremaschi, and
  Vannucci]{argiento2020hierarchical}
Argiento, R., Cremaschi, A., and Vannucci, M. (2020).
\newblock Hierarchical normalized completely random measures to cluster grouped
  data.
\newblock {\em Journal of the American Statistical Association\/}, {\bf 115},
  318--333.

\bibitem[Buckley and James(1979)Buckley and James]{buckley1979linear}
Buckley, J. and James, I. (1979).
\newblock Linear regression with censored data.
\newblock {\em Biometrika\/}, {\bf 66}, 429--436.

\bibitem[Cai {\em et~al.}(2011)Cai, Lin, and Wang]{cai2011bayesian}
Cai, B., Lin, X., and Wang, L. (2011).
\newblock Bayesian proportional hazards model for current status data with
  monotone splines.
\newblock {\em Computational Statistics \& Data Analysis\/}, {\bf 55},
  2644--2651.

\bibitem[Camerlenghi {\em et~al.}(2019)Camerlenghi, Lijoi, Orbanz, and
  Pr{\"u}nster]{camerlenghi2019distribution}
Camerlenghi, F., Lijoi, A., Orbanz, P., and Pr{\"u}nster, I. (2019).
\newblock Distribution theory for hierarchical processes.
\newblock {\em The Annals of Statistics\/}, {\bf 47}, 67--92.

\bibitem[Chipman {\em et~al.}(2010)Chipman, George, McCulloch, {\em
  et~al.}]{chipman2010bart}
Chipman, H.~A., George, E.~I., McCulloch, R.~E., {\em et~al.} (2010).
\newblock {BART}: {B}ayesian additive regression trees.
\newblock {\em The Annals of Applied Statistics\/}, {\bf 4}, 266--298.

\bibitem[Christensen and Johnson(1988)Christensen and
  Johnson]{christensen1988modelling}
Christensen, R. and Johnson, W. (1988).
\newblock Modelling accelerated failure time with a {D}irichlet process.
\newblock {\em Biometrika\/}, {\bf 75}, 693--704.

\bibitem[Cox(1972)Cox]{cox1972regression}
Cox, D.~R. (1972).
\newblock Regression models and life-tables.
\newblock {\em Journal of the Royal Statistical Society: Series B\/}, {\bf 34},
  187--202.

\bibitem[De~Iorio {\em et~al.}(2004)De~Iorio, M{\"u}ller, Rosner, and
  MacEachern]{de2004anova}
De~Iorio, M., M{\"u}ller, P., Rosner, G.~L., and MacEachern, S.~N. (2004).
\newblock An {ANOVA} model for dependent random measures.
\newblock {\em Journal of the American Statistical Association\/}, {\bf 99},
  205--215.

\bibitem[De~Iorio {\em et~al.}(2009)De~Iorio, Johnson, M{\"u}ller, and
  Rosner]{de2009bayesian}
De~Iorio, M., Johnson, W.~O., M{\"u}ller, P., and Rosner, G.~L. (2009).
\newblock Bayesian nonparametric nonproportional hazards survival modeling.
\newblock {\em Biometrics\/}, {\bf 65}, 762--771.

\bibitem[Duan {\em et~al.}(2020)Duan, Young, Nishimura, and
  Dunson]{duan2018bayesian}
Duan, L.~L., Young, A.~L., Nishimura, A., and Dunson, D.~B. (2020).
\newblock Bayesian constraint relaxation.
\newblock {\em Biometrika\/}, {\bf 107}, 191--204.

\bibitem[Dunson and Dinse(2002)Dunson and Dinse]{dunson2002bayesian}
Dunson, D.~B. and Dinse, G.~E. (2002).
\newblock Bayesian models for multivariate current status data with informative
  censoring.
\newblock {\em Biometrics\/}, {\bf 58}, 79--88.

\bibitem[Escobar and West(1995)Escobar and West]{escobar1995bayesian}
Escobar, M.~D. and West, M. (1995).
\newblock Bayesian density estimation and inference using mixtures.
\newblock {\em Journal of the American Statistical Association\/}, {\bf 90},
  577--588.

\bibitem[Ferguson(1973)Ferguson]{ferguson1973bayesian}
Ferguson, T.~S. (1973).
\newblock A {B}ayesian analysis of some nonparametric problems.
\newblock {\em The Annals of Statistics\/}, {\bf 1}, 209--230.

\bibitem[Geweke(1991)Geweke]{geweke1991efficient}
Geweke, J. (1991).
\newblock Efficient simulation from the multivariate normal and student-t
  distributions subject to linear constraints and the evaluation of constraint
  probabilities.
\newblock In {\em Computing Science and Statistics: Proceedings of the
  Twenty-Third Symposium on the Interface\/}, pages 571--578.

\bibitem[Golden {\em et~al.}(2005)Golden, Whittington, Handsfield, Hughes,
  Stamm, Hogben, Clark, Malinski, Helmers, Thomas, {\em
  et~al.}]{golden2005effect}
Golden, M.~R., Whittington, W.~L., Handsfield, H.~H., Hughes, J.~P., Stamm,
  W.~E., Hogben, M., Clark, A., Malinski, C., Helmers, J.~R., Thomas, K.~K.,
  {\em et~al.} (2005).
\newblock Effect of expedited treatment of sex partners on recurrent or
  persistent gonorrhea or chlamydial infection.
\newblock {\em New England Journal of Medicine\/}, {\bf 352}, 676--685.

\bibitem[Groeneboom and Wellner(1992)Groeneboom and
  Wellner]{groeneboom1992information}
Groeneboom, P. and Wellner, J.~A. (1992).
\newblock {\em Information bounds and nonparametric maximum likelihood
  estimation\/}, volume~19.
\newblock Birkhäuser Basel.

\bibitem[Grushka(1972)Grushka]{grushka1972characterization}
Grushka, E. (1972).
\newblock Characterization of exponentially modified {G}aussian peaks in
  chromatography.
\newblock {\em Analytical Chemistry\/}, {\bf 44}, 1733--1738.

\bibitem[Hanson and Johnson(2002)Hanson and Johnson]{hanson2002modeling}
Hanson, T. and Johnson, W.~O. (2002).
\newblock Modeling regression error with a mixture of {P}olya trees.
\newblock {\em Journal of the American Statistical Association\/}, {\bf 97},
  1020--1033.

\bibitem[Hanson and Johnson(2004)Hanson and Johnson]{hanson2004bayesian}
Hanson, T. and Johnson, W.~O. (2004).
\newblock A {B}ayesian semiparametric {AFT} model for interval-censored data.
\newblock {\em Journal of Computational and Graphical Statistics\/}, {\bf 13},
  341--361.

\bibitem[Hjort {\em et~al.}(1990)Hjort {\em et~al.}]{hjort1990nonparametric}
Hjort, N.~L. {\em et~al.} (1990).
\newblock Nonparametric {B}ayes estimators based on beta processes in models
  for life history data.
\newblock {\em The Annals of Statistics\/}, {\bf 18}, 1259--1294.

\bibitem[Ibrahim {\em et~al.}(2001)Ibrahim, Chen, and Sinha]{ibrahim2005b}
Ibrahim, J.~G., Chen, M.-H., and Sinha, D. (2001).
\newblock {\em Bayesian Survival Analysis\/}.
\newblock Springer, New York.

\bibitem[Ishwaran and James(2001)Ishwaran and James]{ishwaran2001gibbs}
Ishwaran, H. and James, L.~F. (2001).
\newblock Gibbs sampling methods for stick-breaking priors.
\newblock {\em Journal of the American Statistical Association\/}, {\bf 96},
  161--173.

\bibitem[James {\em et~al.}(2009)James, Lijoi, and
  Pr{\"u}nster]{james2009posterior}
James, L.~F., Lijoi, A., and Pr{\"u}nster, I. (2009).
\newblock Posterior analysis for normalized random measures with independent
  increments.
\newblock {\em Scandinavian Journal of Statistics\/}, {\bf 36}, 76--97.

\bibitem[Jara {\em et~al.}(2010)Jara, Lesaffre, De~Iorio, and
  Quintana]{jara2010bayesian}
Jara, A., Lesaffre, E., De~Iorio, M., and Quintana, F. (2010).
\newblock Bayesian semiparametric inference for multivariate
  doubly-interval-censored data.
\newblock {\em The Annals of Applied Statistics\/}, {\bf 4}, 2126--2149.

\bibitem[Jara {\em et~al.}(2011)Jara, Hanson, Quintana, M\"uller, and
  Rosner]{dppackage}
Jara, A., Hanson, T., Quintana, F., M\"uller, P., and Rosner, G. (2011).
\newblock {DPpackage}: Bayesian semi- and nonparametric modeling in {R}.
\newblock {\em Journal of Statistical Software\/}, {\bf 40}, 1--30.

\bibitem[Kalbfleisch(1978)Kalbfleisch]{kalbfleisch1978non}
Kalbfleisch, J.~D. (1978).
\newblock Non-parametric {B}ayesian analysis of survival time data.
\newblock {\em Journal of the Royal Statistical Society: Series B\/}, {\bf 40},
  214--221.

\bibitem[Kalli {\em et~al.}(2011)Kalli, Griffin, and Walker]{kalli2011slice}
Kalli, M., Griffin, J.~E., and Walker, S.~G. (2011).
\newblock Slice sampling mixture models.
\newblock {\em Statistics and Computing\/}, {\bf 21}, 93--105.

\bibitem[Kottas and Gelfand(2001)Kottas and Gelfand]{kottas2001bayesian}
Kottas, A. and Gelfand, A.~E. (2001).
\newblock Bayesian semiparametric median regression modeling.
\newblock {\em Journal of the American Statistical Association\/}, {\bf 96},
  1458--1468.

\bibitem[Kuo and Mallick(1997)Kuo and Mallick]{kuo1997bayesian}
Kuo, L. and Mallick, B. (1997).
\newblock Bayesian semiparametric inference for the accelerated failure-time
  model.
\newblock {\em Canadian Journal of Statistics\/}, {\bf 25}, 457--472.

\bibitem[Lo(1984)Lo]{lo1984class}
Lo, A.~Y. (1984).
\newblock On a class of {B}ayesian nonparametric estimates: {I}. {D}ensity
  estimates.
\newblock {\em The Annals of Statistics\/}, {\bf 12}, 351--357.

\bibitem[MacEachern(1999)MacEachern]{maceachern1999dependent}
MacEachern, S.~N. (1999).
\newblock Dependent nonparametric processes.
\newblock In {\em ASA proceedings of the section on Bayesian statistical
  science\/}, volume~1.

\bibitem[MacEachern and M{\"u}ller(1998)MacEachern and
  M{\"u}ller]{maceachern1998estimating}
MacEachern, S.~N. and M{\"u}ller, P. (1998).
\newblock Estimating mixture of {D}irichlet process models.
\newblock {\em Journal of Computational and Graphical Statistics\/}, {\bf 7},
  223--238.

\bibitem[M{\"u}ller {\em et~al.}(2015)M{\"u}ller, Quintana, Jara, and
  Hanson]{muller2015bayesian}
M{\"u}ller, P., Quintana, F.~A., Jara, A., and Hanson, T. (2015).
\newblock {\em Bayesian nonparametric data analysis\/}.
\newblock Springer.

\bibitem[Neal(2000)Neal]{neal2000markov}
Neal, R.~M. (2000).
\newblock Markov chain sampling methods for {D}irichlet process mixture models.
\newblock {\em Journal of Computational and Graphical Statistics\/}, {\bf 9},
  249--265.

\bibitem[Patra and Dunson(2018)Patra and Dunson]{patra2018constrained}
Patra, S. and Dunson, D.~B. (2018).
\newblock Constrained {B}ayesian inference through posterior projections.
\newblock {\em arXiv preprint arXiv:1812.05741\/}.

\bibitem[Quintana {\em et~al.}(2020)Quintana, M{\"u}ller, Jara, and
  MacEachern]{quintana2020dependent}
Quintana, F.~A., M{\"u}ller, P., Jara, A., and MacEachern, S.~N. (2020).
\newblock The dependent {D}irichlet process and related models.
\newblock {\em arXiv preprint arXiv:2007.06129\/}.

\bibitem[Rodriguez {\em et~al.}(2008)Rodriguez, Dunson, and
  Gelfand]{rodriguez2008nested}
Rodriguez, A., Dunson, D.~B., and Gelfand, A.~E. (2008).
\newblock The nested {D}irichlet process.
\newblock {\em Journal of the American Statistical Association\/}, {\bf 103},
  1131--1154.

\bibitem[Sal~y Rosas and Hughes(2011)Sal~y Rosas and
  Hughes]{saly2011nonparametric}
Sal~y Rosas, V.~G. and Hughes, J.~P. (2011).
\newblock Nonparametric and semiparametric analysis of current status data
  subject to outcome misclassification.
\newblock {\em Statistical Communications in Infectious Diseases\/}, {\bf 3},
  364.

\bibitem[Sethuraman(1994)Sethuraman]{sethuraman1994constructive}
Sethuraman, J. (1994).
\newblock A constructive definition of {D}irichlet priors.
\newblock {\em Statistica Sinica\/}, {\bf 4}, 639--650.

\bibitem[Sparapani {\em et~al.}(2016)Sparapani, Logan, McCulloch, and
  Laud]{sparapani2016nonparametric}
Sparapani, R.~A., Logan, B.~R., McCulloch, R.~E., and Laud, P.~W. (2016).
\newblock Nonparametric survival analysis using {B}ayesian additive regression
  trees ({BART}).
\newblock {\em Statistics in Medicine\/}, {\bf 35}, 2741--2753.

\bibitem[Teh {\em et~al.}(2005)Teh, Jordan, Beal, and Blei]{teh2005sharing}
Teh, Y.~W., Jordan, M.~I., Beal, M.~J., and Blei, D.~M. (2005).
\newblock Sharing clusters among related groups: {H}ierarchical {D}irichlet
  processes.
\newblock In {\em Advances in Neural Information Processing Systems\/}, pages
  1385--1392.

\bibitem[Wang {\em et~al.}(2015)Wang, Wang, and McMahan]{wang2015regression}
Wang, N., Wang, L., and McMahan, C.~S. (2015).
\newblock Regression analysis of bivariate current status data under the
  {G}amma-frailty proportional hazards model using the {EM} algorithm.
\newblock {\em Computational Statistics \& Data Analysis\/}, {\bf 83},
  140--150.

\bibitem[Wang and Ding(2000)Wang and Ding]{wang2000assessing}
Wang, W. and Ding, A.~A. (2000).
\newblock On assessing the association for bivariate current status data.
\newblock {\em Biometrika\/}, {\bf 87}, 879--893.

\end{thebibliography}


\clearpage\pagebreak\newpage
\newgeometry{textheight=9in, textwidth=6.5in,}
\pagestyle{fancy}
\fancyhf{}
\rhead{\bfseries\thepage}
\lhead{\bfseries SUPPLEMENTARY MATERIALS}

\baselineskip 20pt
\begin{center}
{\LARGE{Supplementary Materials for\\} 
\bf Bayesian Nonparametric\\
\vskip -8pt
Bivariate Survival Regression\\
\vskip 8pt
for Current Status Data}
\end{center}

\setcounter{equation}{0}
\setcounter{page}{1}
\setcounter{table}{1}
\setcounter{figure}{0}
\setcounter{section}{0}
\numberwithin{table}{section}
\renewcommand{\theequation}{S.\arabic{equation}}
\renewcommand{\thesubsection}{S.\arabic{section}.\arabic{subsection}}
\renewcommand{\thesection}{S.\arabic{section}}
\renewcommand{\thepage}{S.\arabic{page}}
\renewcommand{\thetable}{S.\arabic{table}}
\renewcommand{\thefigure}{S.\arabic{figure}}
\baselineskip=15pt

\begin{center}
Giorgio Paulon$^{1}$ (giorgio.paulon@utexas.edu)\\
Peter M\"uller$^{2}$ (pmueller@math.utexas.edu)\\
Victor G. Sal Y Rosas$^{3}$(vsalyrosas@pucp.edu.pe)\\

\vskip 7mm
$^{1}$Department of Statistics and Data Sciences, \\
University of Texas at Austin,\\ 2317 Speedway D9800, Austin, TX 78712-1823, USA\\
\vskip 8pt 
$^{2}$Department of Mathematics, \\
University of Texas at Austin,\\ 2515 Speedway C1200, Austin, TX 78712-1202, USA\\
\vskip 8pt 
$^{3}$Secci\'{o}n Matem\'{a}ticas, Departamento de Ciencias,\\ 
Pontificia Universidad Cat\'{o}lica del Per\'{u},\\
Av. Universitaria 1801, San Miguel 15088, Peru
\end{center}

\vskip 10mm
Supplementary materials present 
proofs of the theorems illustrated in the main manuscript, 
details of the MCMC algorithm we designed to sample from the posterior, 
convergence diagnostics of the model applied to the real data set
and simulated experiments evaluating the performance of the model framework presented in Section \ref{sec:bivariate} in the main paper.

\newpage

\section{Proofs of Lemma 1 and Theorem 1}
\label{sec:lemma_proof}

\begin{Proof}[Lemma 1]
Recall that 
$$\{C \mid S = s\} = \min \{s + \Exp(\lambda); \Unif(A, B) \}.$$
Then, the inverse cumulative density function for the conditional distribution of censoring times given the latent times is  given by the survival function
\begin{equation*}
    \begin{split}
        \bar{F}_{C \mid S}(c) &= \mathbb{P}[\min \{s + \Exp(\lambda); \Unif(A, B) \} > c]
        \\
        &= \mathbb{P}[s + \Exp(\lambda) > c; \Unif(A, B) > c]
        \\
        &= \mathbb{P}[s + \Exp(\lambda) > c] \cdot \mathbb{P}[\Unif(A, B) > c]
        \\
        &= \frac{B - c}{B - A} \mathds{1}_{(A, B)}(c) \left[ \mathds{1}\{c \leq s\} + e^{-\lambda (c - s)} \mathds{1}\{c > s\} \right].
    \end{split}
\end{equation*}
The condition $\mathds{1}_{(A, B)}(c)$ will be considered to be always true, and hence omitted, in the following. 
This is assured by choosing $A$ and $B$ such that they cover the observation range.
Therefore,
\begin{equation*}
    F_{C \mid S}(c) = \left( 1 - \frac{B - c}{B - A} \right) \mathds{1}\{c \leq s\} + \left( 1 - \frac{B - c}{B - A} e^{-\lambda (c - s)} \right) \mathds{1}\{c > s\},
\end{equation*}
and by differentiation we get
\begin{equation*}
    f_{C \mid S}(c) = \frac{1}{B - A} \mathds{1}\{c \leq s\} + \frac{e^{-\lambda (c - s)}}{B - A}  \{1 + \lambda(B - c)\} \mathds{1}\{c > s\}.
\end{equation*}
\end{Proof}
\qed

\begin{Proof}[Theorem 1]
We begin by calculating the marginal distribution for the infection times as 
\begin{equation*}
    \begin{split}
        f_{I}(I) &= \int f_{I,S}(I,S) dS
        \\
        &= w f_{I}(I) + (1 - w) f_{I}(I) \int f_{L}(S - I) dS
        \\
        &= w f_{I}(I) + (1 - w) f_{I}(I) \int_{I}^{+\infty} \lambda_{L} e^{-\lambda (S - I)} dS = f_{I}(I)
        \\
        &= \sum_{k = 1}^{+\infty} \pi_{k}^{(I)} \mathrm{N} (I \mid \mu_{k}^{(I)}, \sigma_{k}^{(I)2}).
    \end{split}
\end{equation*}

The marginal distribution for the symptoms times is 
\begin{equation*}
    \begin{split}
        f_{S}(S) &= \int f_{I,S}(I,S) dI
        \\
        &= w f_{S}(S) + (1 - w) \int f_{I}(I) f_{L}(S - I) dI
        \\
        &= w f_{S}(S) + (1 - w) \int_{-\infty}^{S} f_{I}(I) \lambda_{L} e^{-\lambda (S - I)} dI
        \\
        &= w \sum_{k = 1}^{+\infty} \pi_{k}^{(S)} \mathrm{N} (S \mid \mu_{k}^{(S)}, \sigma_{k}^{(S)2}) + 
        \\
        &\quad \ (1 - w) \sum_{k = 1}^{+\infty} \pi_{k}^{(I)} \lambda_{L} \exp \left\{ \frac{\lambda_{L}}{2} (\lambda_{L} \sigma_{k}^{2(I)} + 2 \mu_{k}^{(I)} - 2 S) \right\} \Phi \left( \frac{S - \mu_{k}^{(I)} - \lambda_{L} \sigma_{k}^{(I)2}}{\sigma_{k}^{(I)}} \right).
    \end{split}
\end{equation*}
\end{Proof}
\qed

\section{Details of the MCMC Scheme}
\label{sec:mcmc}

\subsection{Prior Hyper-parameters and MCMC Initializations}

The parameters $\lambda$ and $\lambda_{L}$ were assigned Gamma priors $\lambda \sim \text{Ga}(a_{\lambda}, b_{\lambda})$, $\lambda_{L} \sim \text{Ga} (a_{L}, b_{L})$. 
The hyperparameters were chosen to imply the $95\%$ prior credible intervals for the latency times to be $[0.05, 9]$ days, yielding $a_{\lambda} = a_{L} = 10$, $b_{\lambda} = b_{L} = 20$.
The proportion $w$ of individuals with symptoms due to other causes has a $\text{Beta}(a_{w}, b_{w})$ prior. 
The hyperparameters were chosen so that $a_{w} = b_{w} = 1$, i.e. a uniform prior.

The total masses for the two random probability measures $M^{(I)}$ and $M^{(S)}$ are included in the MCMC scheme and assigned Gamma priors, as recommended in \citetlatex{escobar1995bayesian}.
We use $a_{M} = 10, b_{M} = 1$ for both of them. 
Recall the base measures $H_{0}^{(E)}$, $E \in \{I, S\}$,
$$\{ \bm_{k}^{(E)}, \sigma_{k}^{(E)2}\}_{k = 1}^{+\infty} \mytilde{iid} H_0^{(E)} =  \text{N}(\bm_{k}^{(E)} \mid \bm_0^{(E)}, \Sigma_{0}^{(E)}) \times \text{IG}(\sigma_{k}^{(E)2} \mid a_\sigma^{(E)}, b_\sigma^{(E)}).$$
We use vague priors for $m_{0j} \mytilde{iid} \Normal (0, 100^{2})$, $\Sigma_{0}^{(E)} = \text{diag}(\sigma_{0j}^{2})_{j = 1}^{p}$, $\sigma_{0j}^{2} \mytilde{iid} \IG (1, 1)$, $b_{\sigma}^{(E)} \sim \text{Ga}(1, 1)$ ,whereas we fix $a_{\sigma}^{(E)} = 1$.

The initialization for the partitions of infection times and times until symptoms were obtained by using a $K$-means algorithm on the censoring times, with $K = 5$. 
The group-specific location and scale parameters were initialized to the corresponding maximum likelihood estimators. 
The remaining parameters were initialized from their priors.

The algorithm proves to be very robust to both the prior specification and to the initialization.

\subsection{Posterior Computation}
\label{sec:post_alg}

Posterior inference for the bivariate survival regression model, described in Section \ref{sec:bivariate} in the main paper, is based on a posterior Monte Carlo sample generated using a Gibbs sampler simulation.
In what follows, $\bzeta$ denotes a generic variable that collects all other variables not explicitly mentioned, including the data.

The algorithm imputes the latent times to symptoms and times to infection. 
Due to space constraint in the table in Algorithm \ref{algo: MCMC}, we detail here how these parameters can be sampled. 
The times until symptoms due to the infection are sampled from
\begin{equation*}
    \begin{split}
        &p(S_{i} \mid \bzeta) \propto \text{Exp}(\lambda_{L} - \lambda) \big\rvert_{I_{i}}^{C_{i}} \quad \text{if } \Delta_{S_{i}} = 1, r_{i}^{(W)} = 0
        \\
        &p(S_{i} \mid \bzeta) \propto \text{Exp}(\lambda_{L}) \big\rvert_{\max \{C_{i}, I_{i} \}}^{+\infty} \quad \text{if } \Delta_{S_{i}} = 0, r_{i}^{(W)} = 0.
    \end{split}
\end{equation*}
Times until symptoms due to other causes are sampled from
\begin{equation*}
    \begin{split}
        &p(S_{i} \mid r_{i \star}^{(S)} = k, \bzeta) \propto \mathrm{N}(\mu_{k}^{(S)}, \sigma_{k}^{(S) 2})\big\rvert_{-\infty}^{C_{i}} \quad \text{if } \Delta_{S_{i}} = 1, r_{i}^{(W)} = 1
        \\
        &p(S_{i} \mid r_{i \star}^{(S)} = k, \bzeta) \propto \mathrm{N}(\mu_{k}^{(S)}, \sigma_{k}^{(S) 2})\big\rvert_{C_{i}}^{+\infty} \quad \text{if } \Delta_{S_{i}} = 0, r_{i}^{(W)} = 1.
    \end{split}
\end{equation*}
Times until infection are sampled from
\begin{equation*}
    \begin{split}
        &p(I_{i} \mid r_{i}^{(I)} = k, \bzeta) \propto \mathrm{N}(\mu_{k}^{(I)} + \lambda \sigma_{k}^{(I) 2}, \sigma_{k}^{(I) 2})\big\rvert_{-\infty}^{\min \{C_{i}, S_{i}\} } \quad \text{if } \Delta_{I_{i}} = 1, r_{i}^{(W)} = 0
        \\
        &p(I_{i} \mid r_{i}^{(I)} = k, \bzeta) \propto \mathrm{N}(\mu_{k}^{(I)} + \lambda \sigma_{k}^{(I) 2}, \sigma_{k}^{(I) 2})\big\rvert_{C_{i}}^{S_{i}} \qquad \qquad \text{if } \Delta_{I_{i}} = 0, r_{i}^{(W)} = 0
        \\
        &p(I_{i} \mid r_{i}^{(I)} = k, \bzeta) \propto \mathrm{N}(\mu_{k}^{(I)}, \sigma_{k}^{(I) 2})\big\rvert_{-\infty}^{C_{i}} \ \quad \qquad \qquad \qquad \text{if } \Delta_{I_{i}} = 1, r_{i}^{(W)} = 1
        \\
        &p(I_{i} \mid r_{i}^{(I)} = k, \bzeta) \propto \mathrm{N}(\mu_{k}^{(I)}, \sigma_{k}^{(I) 2})\big\rvert_{C_{i}}^{+\infty} \ \quad \qquad \qquad \qquad \text{if } \Delta_{I_{i}} = 0, r_{i}^{(W)} = 1.
    \end{split}
\end{equation*}

As mentioned in the main manuscript, we use a truncated approximation to the infinite mixture model. 
Let $K_{max}$ be the truncation level (in the following, we fix $K_{max} = 40$).
We describe here the case without covariates, although the regression terms are straightforward to include in the algorithm. 
We also do not include the update for the base measure hyperparameters as it consists of a simple normal full conditional.
The sampler for the proposed model of Section \ref{sec:bivariate} comprises the steps outlined in Algorithm \ref{algo: MCMC}.

\begin{algorithm} 
\caption{(Gibbs Sampler)}
\label{algo: MCMC}
\begin{algorithmic}[1]
\vspace{0.2cm}

\Algphase{Updating the symptoms parameters}
\State
For $i = 1, \dots, n$, sample the latent times until symptoms $S_{i}$ as described in Section \ref{sec:post_alg}. 

\State
For $k = 1, \dots, K_{max}$, sample the group specific parameters $\mu^{(S)}_{k}, \sigma^{(S) 2}_{k}$ as 
$$\mu^{(S)}_{k} \mid \bzeta \sim p_{0}(\mu^{(S)}_{k}) \prod_{i \text{ s.t.} r_{i \star}^{(S)} = k} p(S_{i} \mid \mu_{k}^{(S)}, \sigma_{k}^{(S) 2}), \quad \sigma^{(S) 2}_{k} \mid \bzeta \sim p_{0}(\sigma^{(S) 2}_{k}) \prod_{i \text{ s.t.} r_{i \star}^{(S)} = k} p(S_{i} \mid \mu_{k}^{(S)}, \sigma_{k}^{(S) 2}).$$

\State
For $i = 1, \dots, n_{1}$, sample the cluster membership indicators $r_{i \star}^{(S)}$ as 
$$p(r_{i \star}^{(S)} = k \mid \bzeta) \propto \pi_{k}^{(S)} \text{N}(S_{i} \mid \mu_{k}^{(S)}, \sigma_{k}^{(S) 2}).$$

\State
For $k = 1, \dots, K_{max}$, update the weights $\pi^{(S)}_{k} = V_{k}^{(S)} \prod_{\ell < k} (1 - V_{\ell}^{(S)})$, where $V_{k} \mid \bzeta \sim \text{Beta}(1 + n_{k}^{(S)}, M^{(S)} - \sum_{\ell = k+1}^{K_{max}} n_{\ell}^{(S)})$ and $n_{k}^{(S)} = \sum_{i} 1 \{ r_{i \star}^{(S)} = k\}$.

\Algphase{Updating the infection parameters}
\State
For $i = 1, \dots, n$, sample the latent infection times $I_{i}$ as described in Section \ref{sec:post_alg}. 

\State
For $k = 1, \dots, K_{max}$, sample the group specific parameters $\mu^{(I)}_{k}, \sigma^{(I) 2}_{k}$ as 
$$\mu^{(I)}_{k} \mid \bzeta \sim p_{0}(\mu^{(I)}_{k}) \prod_{i \text{ s.t.} r_{i}^{(I)}= k} p(I_{i} \mid \mu_{k}^{(I)}, \sigma_{k}^{(I) 2}), \quad \sigma^{(I) 2}_{k} \mid \bzeta \sim p_{0}(\sigma^{(I) 2}_{k}) \prod_{i \text{ s.t.} r_{i}^{(I)}= k} p(I_{i} \mid \mu_{k}^{(I)}, \sigma_{k}^{(I) 2}).$$

\State
For $i = 1, \dots, n$, sample the cluster membership indicators $r_{i}^{(I)}$ as 
$$p(r_{i}^{(I)} = k \mid \bzeta) \propto \pi_{k}^{(I)} \text{N}(I_{i} \mid \mu_{k}^{(I)}, \sigma_{k}^{(I) 2}).$$

\State
For $k = 1, \dots, K_{max}$, update the weights $\pi^{(I)}_{k} = V_{k}^{(I)} \prod_{\ell < k} (1 - V_{\ell}^{(I)})$, where $V_{k} \mid \bzeta \sim \text{Beta}(1 + n_{k}^{(I)}, M^{(I)} - \sum_{\ell = k+1}^{K_{max}} n_{\ell}^{(I)})$ and $n_{k}^{(I)} = \sum_{i} 1 \{ r_{i}^{(I)} = k\}$.

\Algphase{Updating the global parameters}

\State
Update the dependent censoring parameter $\lambda$ with a M-H transition probability using the target distribution 
$$p(\lambda \mid \bzeta) \propto p_{0}(\lambda) \prod_{i \text{ s.t.} \Delta_{S_{i}} = 1} p (C_{i} \mid S_{i}, \lambda).$$

\State
Sample the indicator for dependent symptoms $r_{i}^{(W)} \mid \bzeta \sim \text{Be} (p_{i})$, where $p_{i} = p_{i,1}^{\star} / (p_{i,0}^{\star} + p_{i,1}^{\star})$, $p_{i,1}^{\star} = w \sum_{k = 1}^{+\infty} \pi_{k}^{(S)} \text{N}(S_{i} \mid \mu_{k}^{(S)}, \sigma_{k}^{(S) 2})$, $p_{i,0}^{\star} = (1 - w) \lambda e^{- \lambda (S_{i} - I_{i})}$.

\State
Sample the probability for dependent symptoms $w \mid \bzeta \sim \text{Beta}(a_{w} + \sum_{i} r_{i}^{(W)}, b_{w} + n - \sum_{i} r_{i}^{(W)})$.

\State
Sample the latency parameter $\lambda_{L} \mid \bzeta \sim \text{Ga} \{ a_{L} + n - \sum_{i} r_{i}^{(W)}, b_{L} + \sum_{i \text{ s.t.} r_{i}^{(W)} = 0} (S_{i} - I_{i}) \}$.

\end{algorithmic}
\end{algorithm}

\subsection{Software, Runtime, etc.}

We programmed in \texttt{R} interfaced with \texttt{C++}. 
A total of $35000$ MCMC iterations were run with the initial $10000$ iterations discarded as burn-in.
The chain was subsequently thinned every $20$ iterations.

The code is available as part of the supplementary materials. 
The MCMC algorithm takes 10 minutes on a Macbook laptop with 8 Gb RAM. 
A `readme' file providing additional implementation details is also included in the
supplementary materials.

\section{Convergence Diagnostics}
\label{sec:conv}

This section presents some MCMC convergence diagnostics for the Gibbs sampler described in Section \ref{sec:mcmc}. 
The results presented here are obtained on the real data analysis.

\begin{figure}[!ht]
	\centering
	\includegraphics[width=.49\linewidth]{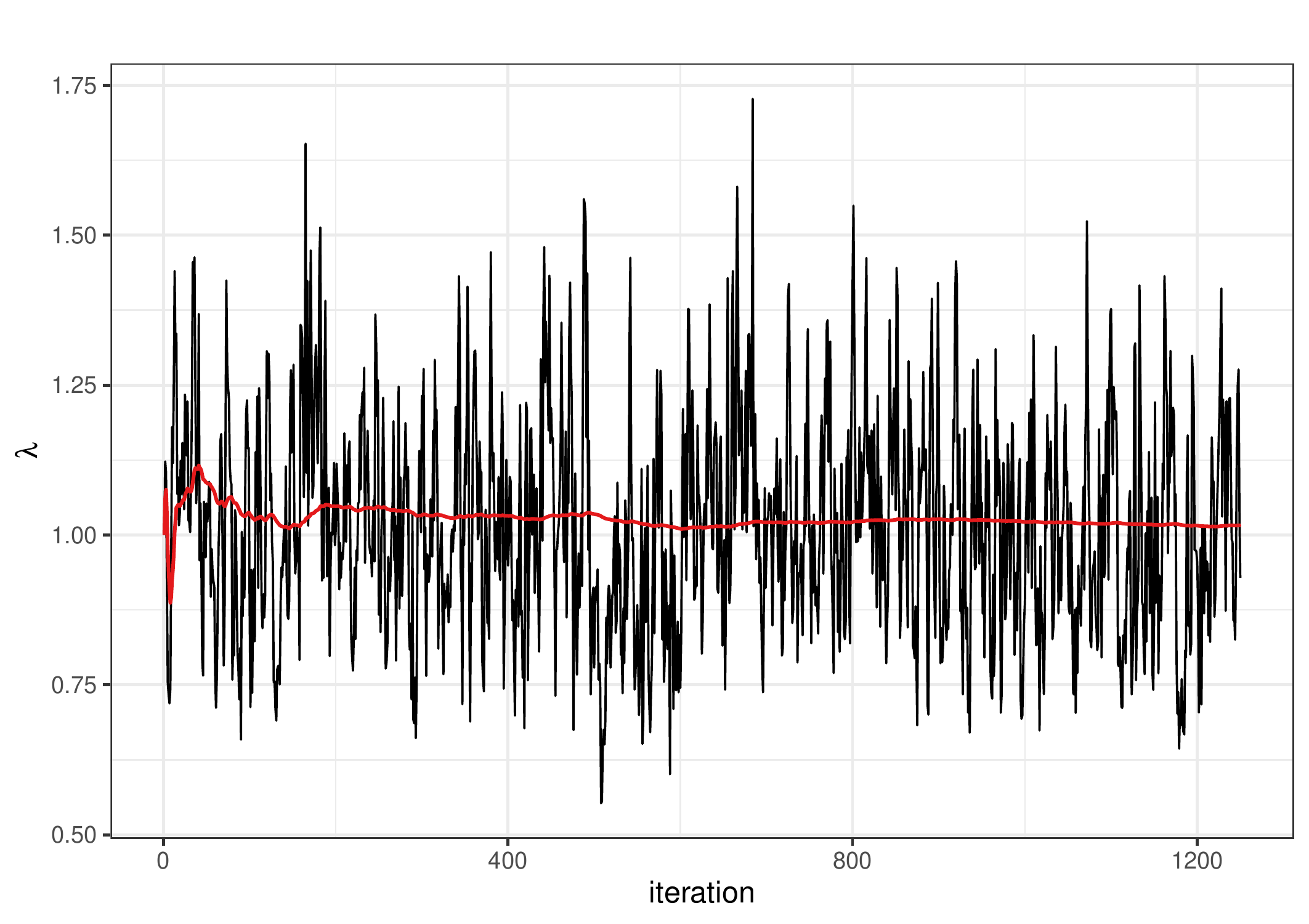}
	\includegraphics[width=.49\linewidth]{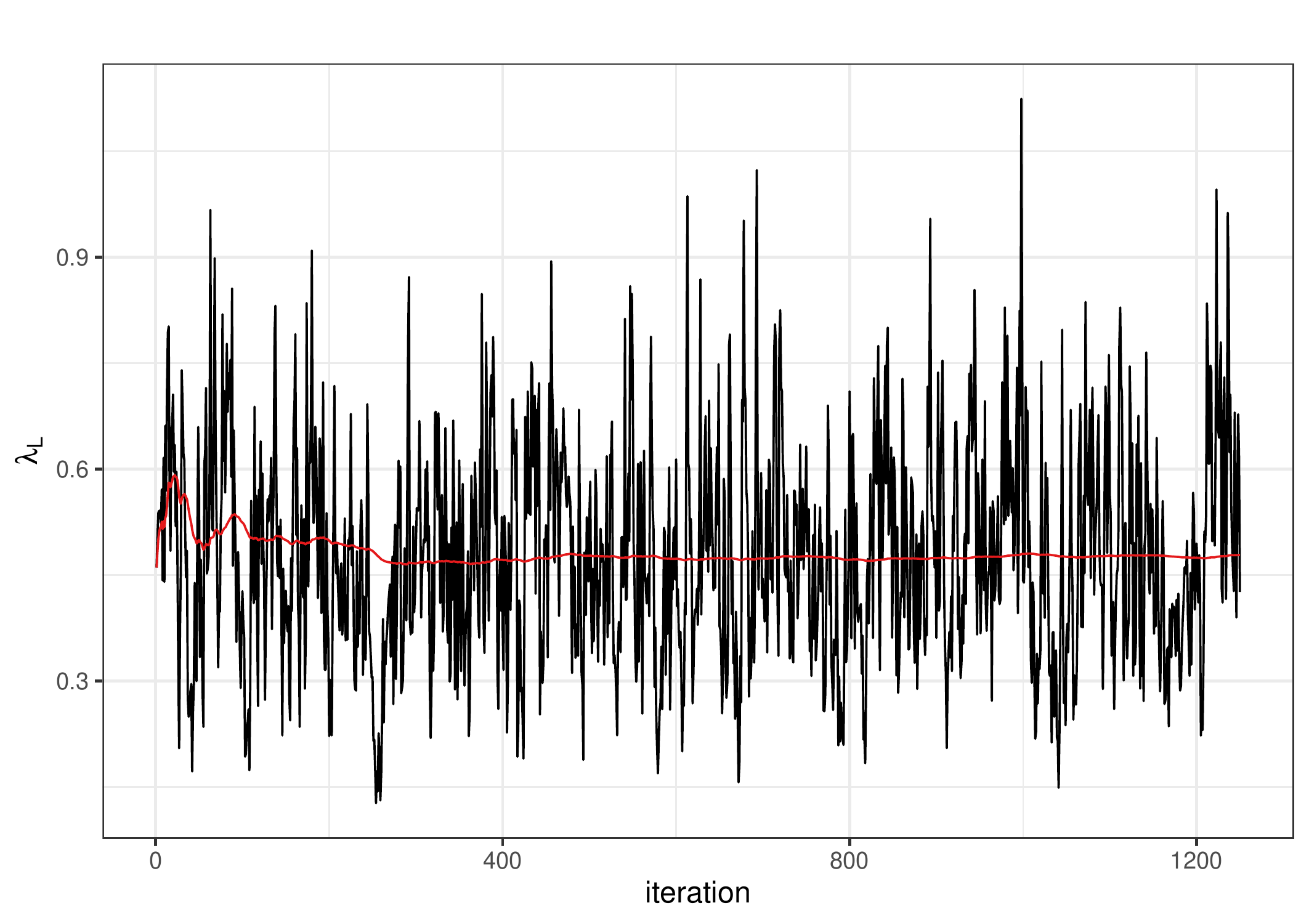}
	\caption{Trace plots of the exponential parameters $\lambda$ and $\lambda_{L}$. 
	In each panel, the solid red line shows the running mean.}
	\label{fig:lambda_trace}
\end{figure}

The Geweke test \citeplatex{geweke1991evaluating} for stationarity of the chains, which formally compares the means of an early vs a later part of a Markov chain (by default the first $10\%$ and the last $50\%$), is also performed. 
If the samples were from the stationary distribution of the chain, the two means are equal and Geweke's statistic has an asymptotically standard normal distribution.
We perform the Geweke test to assess convergence using the global parameters, i.e. those that are not affected by label switching.  
Both the exponential parameters $\lambda$ and $\lambda_{L}$ as well as the proportion of patients with symptoms due to other causes $w$, have very stable traceplots (see Figure \ref{fig:lambda_trace}) and fail to reject the null hypothesis of stationarity of the corresponding chains.

\section{Simulation Studies}
\label{sec:sim_study}

In this section, we discuss the results of some simulated numerical experiments. 
In designing the simulation scenarios, we have tried to closely mimic our recurrent infection dataset. 
We thus chose $n \in \{250, 1000\}$ participants being followed in the time window $[A, B] = [0, 200]$. 
We also simulate two covariates: a \textit{binary} $X_{1}$ and a \textit{continuous} $X_{2}$.
The underlying distribution for the infection times is a mixture of two linear models with weights $\bpi^{(I)} = (0.6, 0.4)^{\intercal}$, location parameters $\bm^{(I)} = [\bm^{(I)}_{1} \quad \bm^{(I)}_{2}] = ((40, -5, 0)^{\intercal}, (100, -10, -15)^{\intercal})$ and scale parameters 

\begin{figure}[!ht]
	\centering
	\includegraphics[width=.49\linewidth]{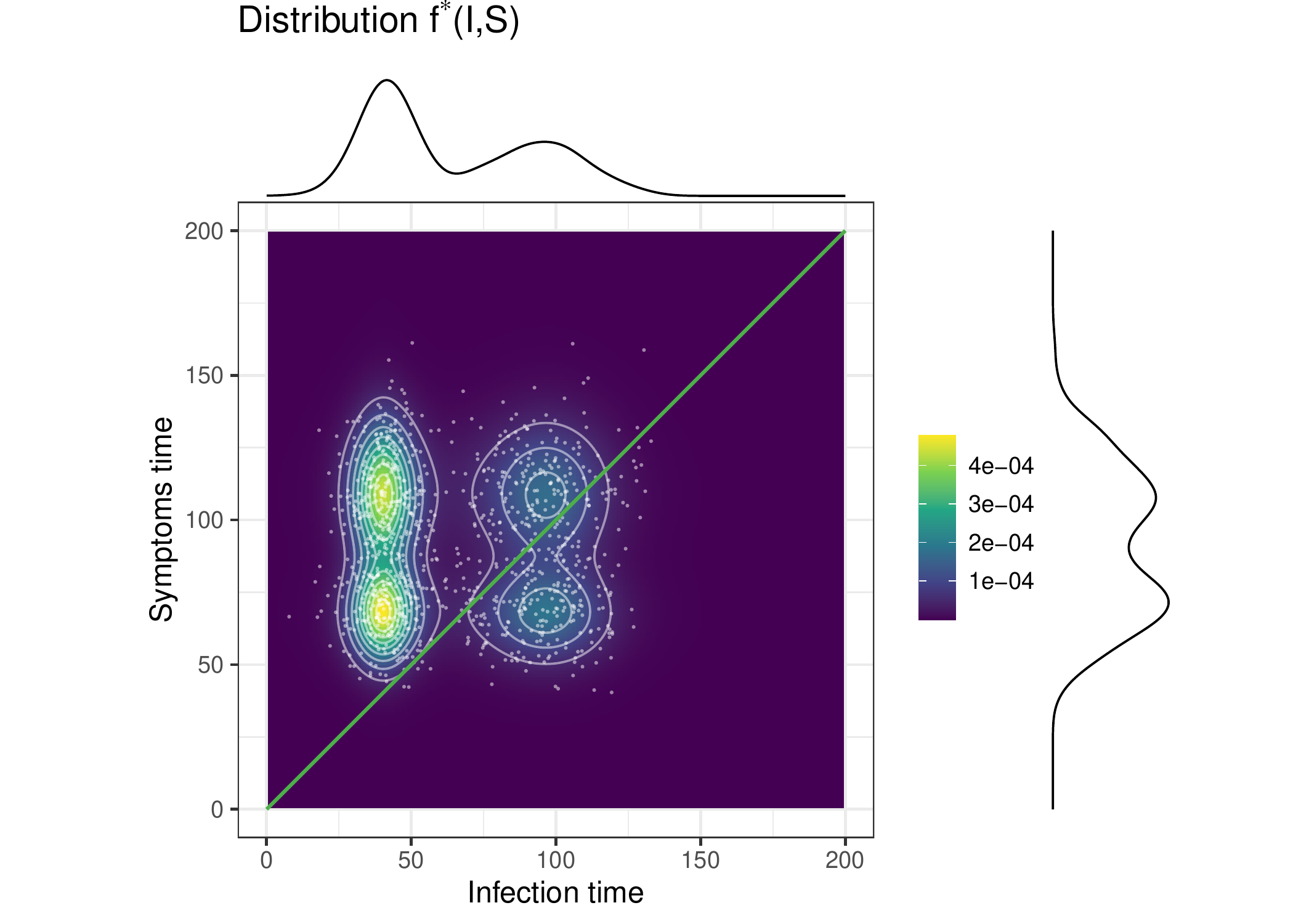}
	\includegraphics[width=.49\linewidth]{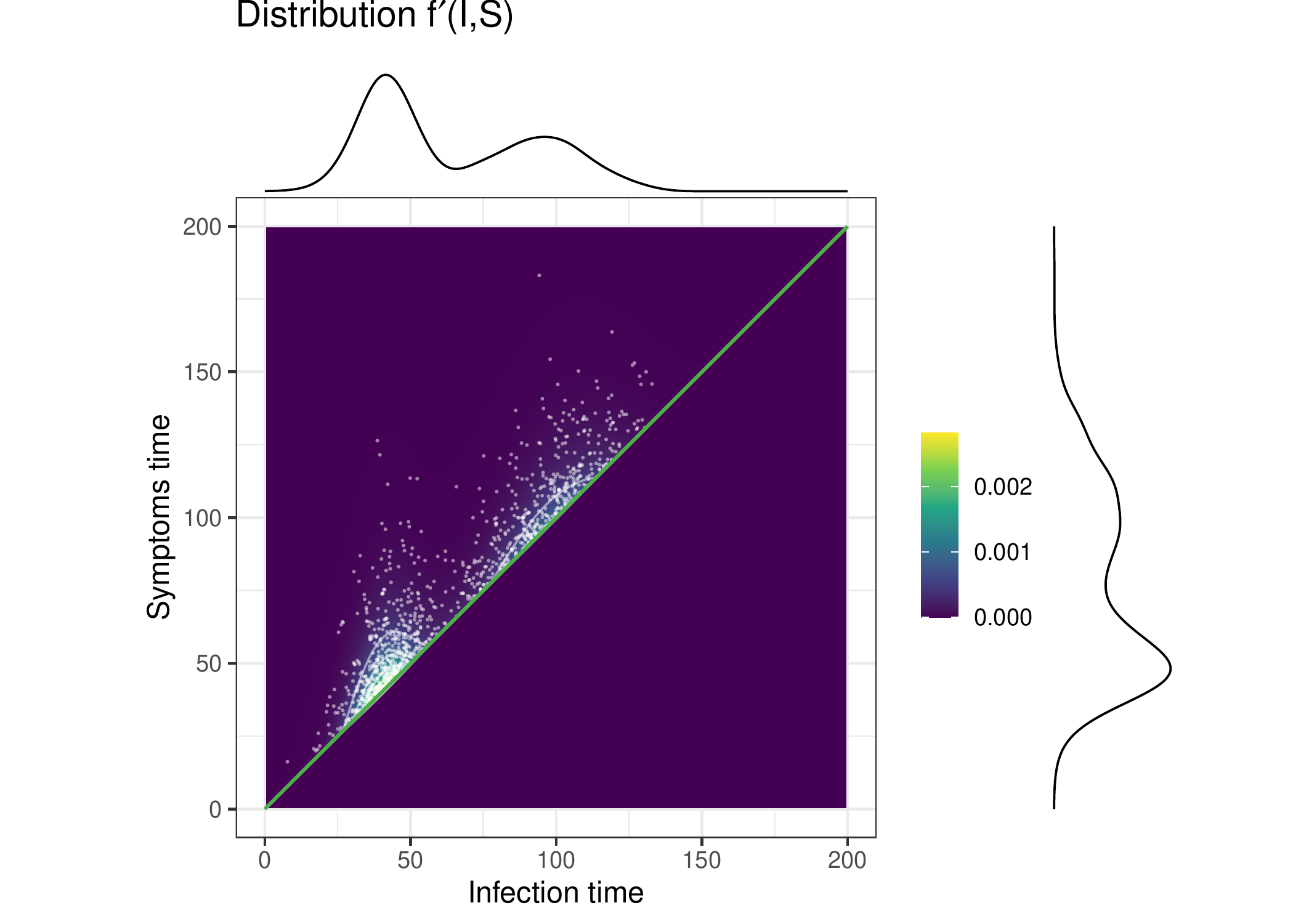}
	\includegraphics[width=.49\linewidth]{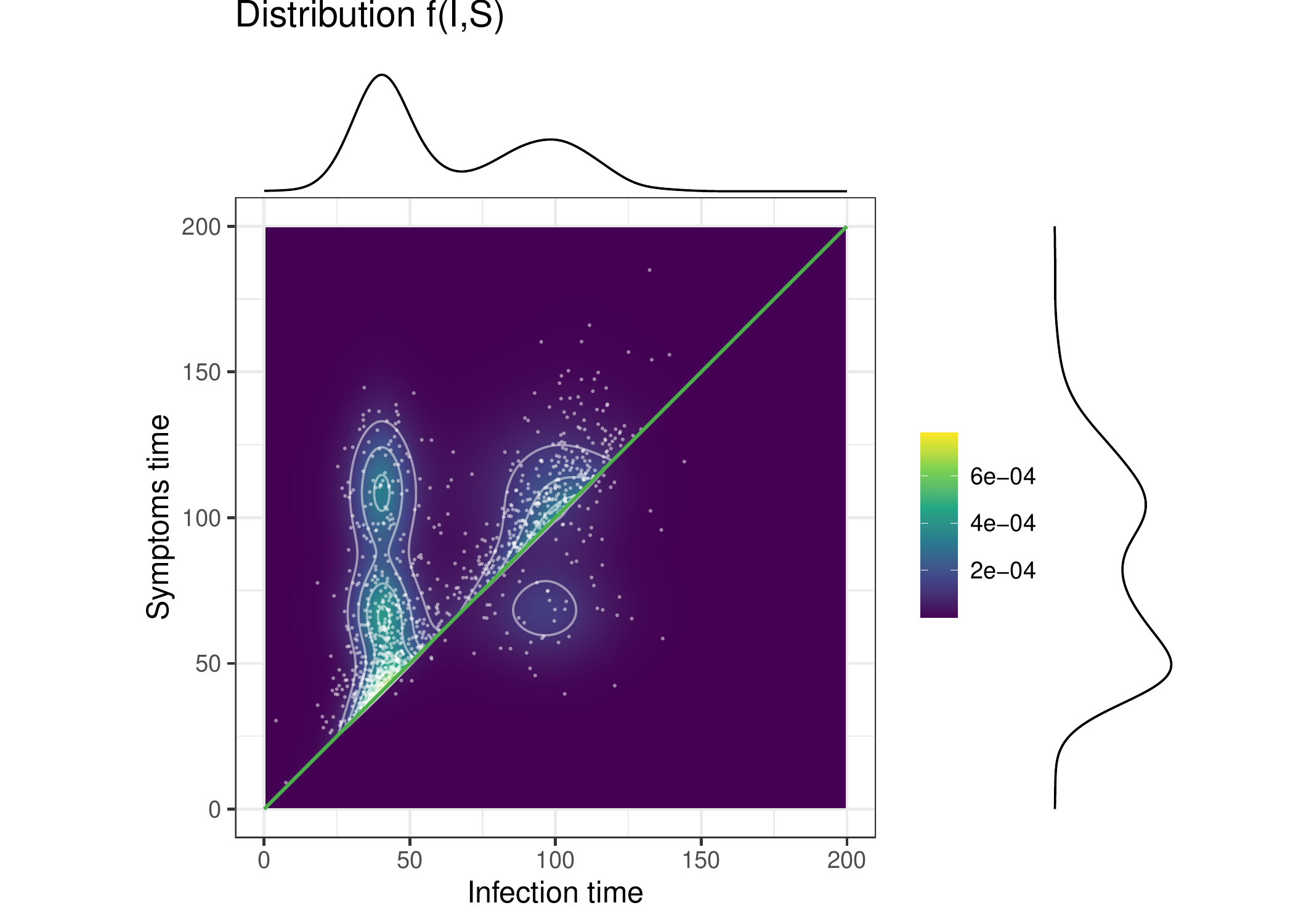}
	\caption{Results for simulated data: Posterior mean density estimate for $f_{IS}^{\star}$, $f_{IS}^{\prime}$ and $f_{IS}$ corresponding to the baseline covariate levels. The green line is the $45^{\circ}$ line $I = S$.
	The corresponding marginal distributions are shown on the top and right side of the density plot. 
	The white points are a sample of the true latent times corresponding to the same covariate levels.}
	\label{fig:sim_results_components}
\end{figure}

$\bsigma^{(I) 2} = (10^{2}, 10^{2})^{\intercal}$.
The distribution for the symptom times due to other causes is a mixture of two linear models with weights $\bpi^{(S)} = (0.4, 0.6)^{\intercal}$, location parameters $\bm^{(S)} = [\bm^{(S)}_{1} \quad \bm^{(S)}_{2}] = ((70, 0, 20)^{\intercal}, (110, -5, 0)^{\intercal})$ and scale parameters 
$\bsigma^{(S) 2} = (10^{2}, 20^{2})^{\intercal}$.
Figure \ref{fig:sim_results_components} shows the results when the proportion of patients with symptoms due to other causes is $w = 0.75$. 
In this simulation, the latency time parameter between infection times and symptom times $\lambda_{L}$ as well as the dependent censoring parameter $\lambda$ are chosen to be $0.2$.
As one can see in Figure \ref{fig:sim_results_components} the underlying true bivariate density is recovered well by our method.
Estimates for other relevant parameters are not reported here, but were also very accurate.

We show how the proposed model compares with two independent ANOVA-DDP models for the marginal distributions in a variety of scenarios. 
In particular, we design three studies when the simulated data have the following features: (I) independent censoring ($\lambda \approx 0$) and dependent symptoms ($w = 0.5$), (II) dependent censoring ($\lambda = 0.2$) and independent symptoms ($w = 1$), and (III) dependent censoring ($\lambda = 0.2$) and dependent symptoms ($w = 0.5$).
All the other parameters are kept fixed as described above.

To evaluate model performance, we measure how well the models are able to recover the functional form of the survival curves for the two marginal distributions.
In particular, we use the mean integrated squared error (MISE). 
The MISE for estimating $f(t)$ by $\hat{f}(t)$ is defined as 
\begin{equation*}
    \mathrm{MISE} = \mathbb{E} \left[ \int \left\{ f(t) - \hat{f}(t) \right\}^{2} dt \right]
\end{equation*}
We estimate the MISE by averaging the estimated integral across $D$ simulated data sets as $\mathrm{MISE_{est}} = \frac{1}{D} \sum_{d=1}^{D} \sum_{i=1}^{N} \Delta_{i} \{ f(t_{i}) - \hat{f}^{(d)}(t_{i}) \}^{2}$, where $\Delta_{i} = t_{i} - t_{i-1}$, $\{t_{i}\}_{i=1}^{N}$ are a set of grid points on the range of the data and $\hat{f}^{(d)}$ is the estimated function of interest for data set $d$. 
In Table \ref{tab:sim_table}, the reported estimated MISEs are based on $D = 50$ simulated data sets.
This simulation shows how the proposed model outperforms the marginal ANOVA-DDP models by exploiting the dependence structure of the data under a wide variety of data generating mechanisms.
\begin{table}[!ht]
    \setlength\arrayrulewidth{0.7pt}
    \renewcommand{\arraystretch}{1.2} 
    \centering
    \begin{tabular}{|c|c|c|c|c|}
        \hline
        Simulation & Sample Size & Distribution & De Iorio et al. & Our method 
        \\ 
        \hline
        \multirow{4}{*}{(I)} & \multirow{2}{*}{$n = 250$} & Inf. & 1.64 (0.92, 3.01) & 1.10 (0.09, 2.24) 
        \\
        & & Sym. & 2.98 (1.11, 5.01) & 1.33 (0.18, 3.72)
        \\ 
        \cline{2-5}
        & \multirow{2}{*}{$n = 1000$} & Inf. & 1.32 (0.73, 1.90) & 0.50 (0.04, 1.80)
        \\
        & & Sym. & 2.32 (1.19, 3.25) & 1.30 (0.54, 2.66)
        \\ 
        \hline
        \multirow{4}{*}{(II)} & \multirow{2}{*}{$n = 250$} & Inf. & 0.96 (0.74, 1.56) & 0.99 (0.13, 2.07) 
        \\
        & & Sym. & 8.44 (5.21, 12.30) & \textbf{0.76} (0.22, 2.16)
        \\ 
        \cline{2-5} 
        & \multirow{2}{*}{$n = 1000$} & Inf. & 0.80 (0.50, 1.10) & \textbf{0.19} (0.05, 0.50) 
        \\
        & & Sym. & 8.18 (6.28, 10.32) & \textbf{0.12} (0.02, 0.37)
        \\ 
        \hline
        \multirow{4}{*}{(III)} & \multirow{2}{*}{$n = 250$} & Inf. & 4.45 (3.00, 6.30) & \textbf{0.45} (0.08, 1.14) 
        \\
        & & Sym. & 9.82 (6.70, 13.20) & \textbf{0.24} (0.03, 0.81)
        \\ 
        \cline{2-5} 
        & \multirow{2}{*}{$n = 1000$} & Inf. & 4.10 (3.18, 4.96) & \textbf{0.13} (0.01, 0.35) 
        \\
        & & Sym. & 9.94 (8.44, 11.71) & \textbf{0.05} (0.01, 0.15)
        \\ 
        \hline
    \end{tabular}
    \caption{Results for simulated data: Estimated median integrated squared error ($\mathrm{MISE_{est}}$) performance of the survival regression model described in Section \ref{sec:bivariate} compared with the method of De Iorio et al. (2009). 
    We have reported here the MISE values for estimating the two marginal distributions (infection and symptoms, respectively) corresponding to the baseline covariate levels. 
    In parenthesis, the $95\%$ credible intervals for the MISE values are reported.
    When a method significantly outperforms the other, the corresponding MISE value is highlighted in bold.}
    \label{tab:sim_table}
\end{table}

\clearpage
\bibliographystylelatex{natbib}
\bibliographylatex{biblio}

\end{document}